%% file: Preprint.tex
\documentclass[preprint,journal]{IEEEtran} 
\usepackage{standalone}

\usepackage{cite}

\usepackage{graphicx}
\usepackage{epstopdf}
\usepackage[subrefformat=parens,labelformat=parens]{subfig}

\graphicspath{{Figs/}}
\DeclareGraphicsExtensions{.eps}
\ifCLASSINFOpdf
\else
\fi
\usepackage{amsmath}
\usepackage{amssymb}
\usepackage{amsthm}
\usepackage{bigints}
\usepackage{algorithm}
\usepackage{algcompatible}
\usepackage{relsize}

\usepackage{tikz}
\usetikzlibrary{shapes, patterns}

\usepackage{array}

\usepackage{amsthm}

\theoremstyle{plain}
\newtheorem{thm}{\protect\theoremname}
\providecommand{\theoremname}{Theorem}

\theoremstyle{plain}
\newtheorem{prop}{\protect\propositionname}
\providecommand{\propositionname}{Proposition}

\theoremstyle{plain}

\newtheorem{corollary}{Corollary}

\newcommand{\A}{\mathcal{A}}
\newcommand{\N}{\mathcal{N}}

\newcommand{\X}{\mathbf{X}}
\newcommand{\x}{\mathbf{x}}

\newcommand{\Vt}{\widetilde{V}}
\newcommand{\Pt}{\widetilde{P}}

\newcommand{\zt}{\widetilde{z}}

\newcommand{\sigmat}{\widetilde{\sigma}}
\newcommand{\SumM}{\sum\limits^M_{m=1}}

\newcommand{\C}{\ensuremath{\mathcal{C}}}
\newcommand{\Hyp}{\ensuremath{\mathcal{H}}}

\newcommand{\LLLR}{\Lambda_{\text{{\scriptsize LLR}}}}

\newcommand{\E}{\ensuremath{\mathbb{E}}}
\newcommand{\Prob}{\ensuremath{\mathbb{P}}}

\newcommand{\var}{\text{var}}
\newcommand{\Pois}{\text{Pois}}

\newcommand{\Xixm}{\left\Vert \Xbi - \x_m \right\Vert^{\frac{\alpha}{2}}}
\newcommand{\Xixmsq}{\left\Vert \Xbi - \x_m \right\Vert^{\alpha}}
\newcommand{\muhmj}[1]{\hat{\mu}_{m,#1}}
\newcommand{\sighmj}[1]{\hat{\sigma}_{m,#1}}
\newcommand{\sighSqmj}[1]{\hat{\sigma}^2_{m,#1}}
\newcommand{\mubmj}{\bar{\mu}_{m,j}}
\newcommand{\sigbmj}{\bar{\sigma}^2_{m,j}}
\newcommand{\Imu}[1]{I_{\bar{\mu}_{m,#1}}}
\newcommand{\Isig}[1]{I_{\bar{\sigma}^2_{m,#1}}}
\newcommand{\sumCm}{\sum_{\X_i \in \Phi_m}}
\newcommand{\SigmaAM}[1]{\hat{\sigma}^2_{m,#1}}

\newcommand{\sumPhim}{\sum_{\Xbi\in\Phi_m}}
\newcommand{\SNRchm}{\text{SNR}^{\text{ch}}_m}

\newcommand{\dB}{\text{dB}}

\def\f2G{f^2\left(G\right)}

\def\Xbi{\mathbf{X}_{i}}

\def\xb{\mathbf{x}}

\def\sumPhi{\sum_{\Xbi\in\Phi}}

\def\Xixm{\left\Vert \Xbi-\xb_{m}\right\Vert ^{\frac{\alpha}{2}}}

\def\Xix0{\left\Vert \Xbi-\xb_{0}\right\Vert}

\def\Xi0{\left\Vert \Xbi \right\Vert}

\def\Xialpha2{\left\Vert \Xbi \right\Vert^{\frac{\alpha}{2}}}

\def\I{I\left(\Xbi\right)}

\def\R{{\mathbb{R}}} 

\def\H{{\cal H}}

\def\C{{\cal C}}

\def\E{\mathbb{E}}

\def\var{\text{var}}

\def\Pdx{P_{d}\left(\xb,\xb_{t}\right) }

\usepackage{comment}
\specialcomment{new}{\begingroup\color{blue}}{\endgroup}
\specialcomment{warning}{\begingroup\color{red}}{\endgroup}
\specialcomment{OK}{\begingroup\color{black}}{\endgroup}

\begin{document}
%

%
\title{Distributed Detection Fusion in Clustered Sensor Networks over Multiple Access Fading Channels}
%
%
%
\author{Sami A. Aldalahmeh,~\IEEEmembership{Senior Member,~IEEE} and Domenico Ciuonzo,~\IEEEmembership{Senior Member,~IEEE}
}

\maketitle

\input{1_abstract_rev2}

\IEEEpeerreviewmaketitle

\input{2_introduction_rev2}

\input{3_related_work_rev2}

\input{4_system_model_rev3}

\input{5_dist_trans_comb_rev2}
\input{6_dist_det_multiple_clusters_rev3}
\input{7_Performance_analysis_rev3}
\input{8_simulation_results_rev3}
\input{9_conclusions_rev2}


\input{9b_Appendices_rev5}

%
%
%
%


\ifCLASSOPTIONcaptionsoff
  \newpage
\fi



%
\bibliographystyle{IEEEtran}
\bibliography{Main_Database}
\vspace{-33pt}

\end{document}

%% file: 1_abstract_rev2.tex
\begin{abstract}
In this paper, we tackle decision fusion for distributed detection in a randomly-deployed clustered Wireless Sensor Networks (WSNs) operating over a non-ideal multiple access channels (MACs), i.e. considering Rayleigh fading, path loss and additive noise.
To mitigate fading, we propose the distributed equal gain transmit combining (dEGTC) and distributed maximum ratio transit combining (dMRTC).
The first and second order statistics of the received signals were analytically computed via stochastic geometry tools. Then the distribution of the received signal over the MAC are approximated by Gaussian and log-normal distributions via moment matching. This enabled the derivation of moment matching optimal fusion rules (MOR) for both distributions. Moreover, suboptimal simpler fusion rules were also proposed, in which all the CHs data are equally weighed, which is termed moment matching equal gain fusion rule (MER). It is shown by simulations that increasing the number of clusters improve the performance. Moreover, MOR-Gaussian based algorithms are better under free-space propagation whereas their lognormal counterparts are more suited  in the ground-reflection case. Also, the latter algorithms show better results in low SNR and SN numbers conditions.
We have proved that the received power at the CH in MAC is proportional $\mathcal{O}\left(\lambda^2 R^2 \right) $ and to $\mathcal{O}\left(\lambda^2 \ln^2 R \right) $ in the free-space propagation and the ground-reflection cases respectively, where $\lambda$ is SN deployment intensity and $R$ is the cluster radius. This implies that having more clusters decreases the required transmission power for a given SNR at the receiver.
\end{abstract}


%% file: 2_introduction_rev2.tex
\section{Introduction}

\IEEEPARstart{W}{ireless sensor networks} (WSNs) are becoming a mainstream technology constituting the backbone of several emerging technologies, such as the Internet of Things (IoT)~\cite{Khalil2014} and smart cities~\cite{Rashid2016} (see references therein).
Indeed, the flexible nature of WSNs~\cite{Chong2003} enables them to pervade such a wide spectrum of applications.
However, several methodological aspects of WSNs remain fertile research grounds, especially those concerning distributed detection (DD)~\cite{Chamberland2007}.
\begin{OK}
In such a scenario, battery-powered sensor nodes (SNs) may be geographically distributed in a vast region of interest (ROI) to monitor it and detect the unexpected presence of 
an intruder (or the occurrence of an anomalous phenomenon of interest, such as an oil leak or a forest fire \cite{alkhatib2017smart}).
Our work thus focuses on surveillance/anomaly detection applications where the simultaneous presence of multiple intruders (viz. anomalous phenomenons of interest) is a somewhat rarer scenario.
\end{OK}

\begin{OK}
The locations of the SNs are best modeled as a \emph{random point process}~\cite{Zhang2015}, since they might be out of communication range, out of power, or/and be randomly deployed (e.g. might be even dropped from an airplane to form a network~\cite{Song2009}).
\end{OK}
Due to constrained power and bandwidth, the collected data is often compressed into a single bit decision. Moreover, the limited  SNs communication range renders providing ubiquitous coverage in large WSNs a challenging task.
Accordingly, the WSN is usually divided into geographical clusters~\cite{Bandyopadhyay2003} and organized hierarchically into three tiers;
($i$) SNs, ($ii$) cluster heads (CHs) and ($iii$) the fusion center (FC). 
The SNs in each cluster send their data to the CH, which usually has access to larger power resources and is able to provide a larger communication range. The CHs in turn report the collected data to the FC, thus acting as moderate-power relays.
Such data is often relayed over imperfect communication channels in either an amplify-and-forward (AF) or decode-and-forward (DF) fashion  \cite{Hong2007}.

In this paper, we investigate the decision fusion for distributed detection in a randomly deployed clustered-WSN operating over nonideal multiple access channels (MACs).
We build on the framework proposed for detection in clustered WSN~\cite{Aldalahmeh2019} and generalize it by considering Rayleigh fading, path loss, and additive noise presence in the channels between SNs and CHs (termed SN-CH).
Also, the channels between CHs and the FC (termed CH-FC) are assumed to suffer additive noise, since the CH is assumed to have more capabilities. 
To the best of the authors' knowledge, this is the first work that studies fusion rules in the above problem setting.

In the light of the previous framework, the main contributions of this paper are:
\begin{enumerate}
\item We propose two distributed transmit combining schemes; distributed equal gain transmit combining (dEGTC) and distributed maximum ratio transit combining (dMRTC), in order to mitigate fading. Interestingly, it is shown that the dEGTC performs better than the dMRTC.
\item
The statistics of the received signals at the CHs are computed via \emph{stochastic-geometry} tools. 
Consequently, Gaussian and lognormal distributions are used to approximate the received signal distribution using moment matching.

\item 
We derive the optimal fusion rule in the Neyman-Pearson sense for both the Gaussian and lognormal cases. Also, we propose a simpler suboptimal fusion rule, which ultimately performs as good as the optimal one when the number of SNs increases.
 
\item
We prove that in the MAC network case, the received power at the CH increases as the networks expands. In fact, the received power increases proportionally to $\mathcal{O}\left(\lambda^2 R^2 \right) $ and  $\mathcal{O}\left(\lambda^2 \ln^2 R \right) $ in the free-space propagation and the ground-reflection cases respectively, where $\lambda$ is SN deployment intensity and $R$ is the cluster radius. This starkly contrasts the parallel access channel (PAC) case, where the received power at the CH decreases as the the SN-CH distance increases.

\end{enumerate}

We highlight the present study extends previous conference work in~\cite{Aldalahmeh2019a}, and includes both optimal and suboptimal fusion rules design in addition to addressing the transmission power issue in the WSN. 

The rest of the paper is organized as follows. 
In Sec.~\ref{sec:Related-Work} related work is reviewed. The system model is presented in Sec.~\ref{sec:System-Model}. The distributed transmit combining techniques and the CHs' received signal statistics are  discussed in Sec.~\ref{sec:Dist-Trans-Comb}, whereas corresponding fusion rules for the multiple cluster case are investigated in Sec.~\ref{sec:Multiple-Clusters}. Section \ref{sec:Perf-analys} provides an insight into the detection performance and  received power analysis for the previous fusion rules. Section~\ref{sec:Simulation-Result} presents the simulation results and their discussions.
Conclusions are drawn in Sec.~\ref{sec:conlcusions}, in addition to a brief discussion of future research direction.

%% file: 3_related_work_rev2.tex
\section{Related Work}\label{sec:Related-Work}
\begin{OK}
DD has been extensively investigated for various sensor network forms, such as parallel, tandem and tree structures~\cite{Viswanathan1997, Blum1997, Tay2008, Tay2009}, and even decentralized architectures~\cite{maya2021fully}.
\end{OK}
\begin{OK}
DD over multiple access channel was also investigated in~\cite{Chamberland2003,Li2007, Liu2007} from the information theoretic and rule design~\cite{jamoos2020distributed} aspects, respectively.
\end{OK}
Whereas type-based DD in MAC context was considered in~\cite{Anandkumar2007}~and~\cite{Liu2007a}.
The Rao test and its generalized version were investigated in~\cite{Ciuonzo2013,ciuonzo2020bandwidth,Ciuonzo2017a,Ciuonzo2017} where a trade-off between complexity and performance has been shown.

However, the previous detectors generally suffered from spurious detection problem
\begin{OK}
\footnote{In which SNs far from the target falsely detect it, due to the sensing signal attenuation, and hence causes performance degradation \cite{Guerriero2009}}.
\end{OK}
This problem might be handled via scan statistics-based detection~\cite{Guerriero2010} and local vote decision fusion rule (LVDF)~\cite{Katenka2008} but at the expense of a significant communication and delay.

\begin{OK}
Clustering~\cite{Abbasi2007} in sensor networks provides an efficient solution to spurious detection and an improved means to parsimonious estimation~\cite{shirazi2020distributed}.
\end{OK}
Hard decision for clustered WSN over multiple-hop binary symmetric channel was investigated in~\cite{Tian2007}, where an optimal fusion rules were derived, however requiring the knowledge of the decision error probability in each sensor in addition to the and bit error probability in each channel. In~\cite{Ferrari2011} majority-like fusion (MLF) rules were implemented in both the CH and FC levels, where surprisingly it was shown that clustering decreases the detection performance. Detection performance was investigated in~\cite{Yemini2020} for intermittent communication between the sensor clusters and the FC residing in the clouds. 
The optimal-cluster-based fusion rule (OCR) for clustered sensor networks was presented in~\cite{Aldalahmeh2016}, where the communication channels were ideal. This scenario was extended to noisy channels in our previous work~\cite{Aldalahmeh2019}. Optimal fusion rules were proposed in addition to an optimal power allocation strategy for the CHs transmission. Fading channels effect on DD in clustered WSN was investigated in~\cite{Eritmen2014}, where channel side information (CSI) was instrumented to derive the optimal detector. 

In the context of the previous literature, this paper, as stated earlier, considers optimal and suboptimal fusion rules for DD in the case of clustered WSNs suffering from channel noise, path-loss and fading.

%% file: 4_system_model_rev3.tex
\section{System Model}\label{sec:System-Model}
\begin{figure}[!ht]
\centering
\scalebox{0.45}{\input{WSN_Fig_Mod.tex}}
\caption{The WSN topology, in which the star is the target, gray-shaded nodes are the detecting SNs and white-shaded nodes are the non-detecting SNs.}
\label{fig:WSN_fig}
\end{figure}
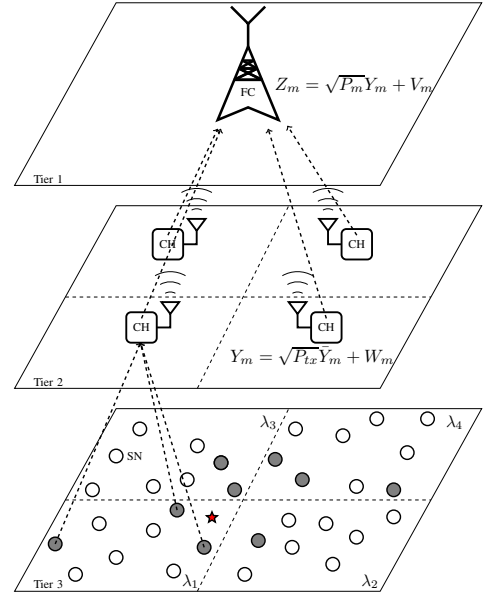

The considered WSN architecture is functionally divided into \emph{three tiers}, as shown in Fig.\:\ref{fig:WSN_fig}, where: tier 1 contains the FC; tier 2 contains the CHs (which are connected to the FC via dedicated channels); tier 3 contains the SNs in the clusters. Note that the SNs in each cluster communicate with the corresponding CH over a shared channel.
In this section, we present: ($a$) the stochastic geometry model for the SNs deployment (similar to \cite{Niu2005,Zhang2015}) and the corresponding sensing model; ($b$) the communication model between the three tiers.
In this work the SNs are assumed to be restricted in both power and bandwidth. On the other hand, the CHs are assumed to have access to higher power and larger bandwidth.

Hereinafter we generally refer to deterministic values by lowercase symbols, bold  symbols refer  to  vector  values, whereas random  values  are  referred  to  by  uppercase  symbols. 
For example, $Y_m$ and $Z_m$ are RVs whereas $y_m$ and $z_m$ are their corresponding realizations. 
Table \ref{tab:Not-sym} collects the notation and most common used variables throughout the paper.

%

\begin{table}[]
    \centering
    \begin{tabular}{|c|l|}
        \hline
        \textbf{Symbol} & \textbf{Meaning} \\
        \hline
        $\Vert \cdot \Vert$  & Euclidean distance\rule{0pt}{2.6ex}  \\
        $\vert \cdot \vert$ & Modulus \\
        $(\cdot)^*$ & Conjugate \\
        $\Prob(A)$ & Probability of event $A$ \\
        $\E[\cdot], \E_X[\cdot]$ & Expectation and expectation w.r.t $X$ \\
        $\var_X(\cdot)$ & Variance w.r.t $X$ \\
        $\N(\mu, \sigma^2)$ & Normal pdf with mean $\mu$ and variance $\sigma^2$  \\
        $\mathcal{Q}(\cdot)$ & Q-function \\
        $\Pois(\rho)$ & Poisson pdf with mean $\rho$\rule[-0.9ex]{0pt}{0pt} \\
        \hline\hline
        $\Phi$ & Overall Poisson point process PPP\rule{0pt}{2.6ex}  \\
        $\lambda$ & Intensity of $\Phi$\\
        $ \Phi_m$ & PPP in the $m$th cluster \\
        $\lambda_m$ & Intensity of $\Phi_m$\\
        $M$ & Number of clusters \\
        $\A$ & Sensing field\\
        $\C_m$ & $m$th cluster zone\\
        $\x_i$ & $i$th SN Cartesian location \\
        $P_{fa}$ & Local SN false alarm probability\\
        $P_d(\x_i)$ & Local SN detection probability at location $\x_i$\\
        $P_{FA}$ & Global probability of false alarm \\
        $P_D$ & Global probability of detection \\
        $P_t$ & Target's signal power \\
        $P_{tx}$ & SN Tx power \\
        $P_m$ & $m$th CH Tx power\\
        $\Pt_m$ & $m$th channel aggregate Tx power \\
        $\alpha$ & Comm. channel path-loss exponent \\
        $H_{m,i}$ & Complex channel gain of the $i$th SN in the $m$th cluster\\
        $\sigma^2_s$ & Sensing noise variance \\
        $\sigma^2_{c,m}$ & $m$th SN-CH channel noise variance \\
        $\sigma^2_{f,m}$ & $m$th CH-FC channel noise variance \\
        $\textrm{SNR}^{\textrm{ch}}_m$ & SNR of the $m$th SN-CH channel \\
        $\textrm{SNR}^{\textrm{fc}}_m$ & SNR of the $m$th CH-FC channel \\
        $\bar{Y}_m$ & $m$th CH noiseless received signal \\
        $Y_m$ & $m$th CH received signal \\
        $Z_m$ & FC received signal from the $m$th CH \\
        $\bar{\mu}_{m,j}, \bar{\sigma}^2_{m,j}$  & Mean and variance of $\bar{Y}_m$ under $\Hyp_j$\\
        $\mu_{m,j}, \sigma^2_{m,j}$  & Mean and variance of $Z_m$ under $\Hyp_j$\rule[-0.9ex]{0pt}{0pt}\\
        \hline
    \end{tabular}
    \caption{Notation and most commonly used symbols.}
    \label{tab:Not-sym}
\end{table}

\subsection{SNs Deployment and Sensing Models} \label{subsec:sensing-model}
Consider a WSN randomly-deployed over a region, $\A \subset \mathbb{R}^2$ where $\A$ is assumed to be significantly large. 
The WSN is modeled by a Poisson point process (PPP) $\Phi = \{ \X_1, \X_2, \cdots, \X_N \}$ in $\A$ \cite{Streit2010}, where $\X_i \in \Phi$ is the coordinate of the $i$th SN.
\begin{OK}
PPPs have been successfully employed to accurately model WSN random deployments in DD tasks ~\cite{Guerriero2010,Zhang2015,Aldalahmeh}.
\end{OK}
This implies that the $\X_i$'s are random variables (RVs) and their number $N = \vert \Phi \vert$ is a Poisson RV having the distribution $N \sim \Pois(\E\left[N\right])$ where $\E\left[N\right]$ is the average number of SNs.
In general, the PPP intensity (the average number of SNs in a unit area) might be non-homogeneous, i.e., the intensity $\lambda(\x)$ is location dependent.
This case might arise due to environmental or application specific constraints.

The WSN is tasked with the detection of any intruder (viz. target) entering the ROI. 
A target at location $\mathbf{x}_t \in \A, \mathbf{x}_t \notin \Phi$  leaves a signature signal sensed by the SNs, which might be thermal, magnetic, electrical, seismic or  electromagnetic signal \cite{Arora2004}.
\begin{OK}
We adopt the sensing model in \cite{Niu2008}, in which the signature power in the far-field is assumed to follow the inverse-square law.%
\footnote{We highlight that our study virtually applies to any general sensing model.}
\end{OK}
The target's parameters are given in the vector $\boldsymbol{\theta} = [P_t, \mathbf{x}_t]^T$, where $P_t$ is the target's signal power. 
The \emph{noise-free} signal received at the $i$th SN located at a given $\x_i$ has the following amplitude:
\begin{align}
a(\x_{i},\boldsymbol{\theta})= & \sqrt{P_{t}}\,/\,\max\left(d_{0},d_{i}\right)\label{eq:noise-free-signal}
\end{align}
where $d_0$ is the reference distance to the node's sensor and $d_i = \| \x_t - \x_i \|$ is the distance between the target and the $i$th SN. Note that the measured signal is saturated if the distance to the target is smaller than $d_0$. The above model can adequately describe acoustic or electromagnetic signals. 

For a given realization of $\Phi$, each SN samples the environment to decide whether an intruder is present or not.
Hence, the collected data $S(\cdot)$ at the $i$th SN under the null ($\mathcal{H}_0$) and alternative ($\mathcal{H}_1$) hypotheses takes the following form:
\begin{equation}
\begin{cases}
\mathcal{H}_{0}\,:\, & S(\x_i)\,=\,Q_{i}\\
\mathcal{H}_{1}\,:\, & S(\x_i)\,=\,a(\x_{i},\boldsymbol{\theta})+Q_{i}
\end{cases}
\label{eq:hypothesis testing problem}
\end{equation}
where $Q_i\sim \mathcal{N}(0,\sigma^2_s)$. 
The noise is assumed to be independently and identically distributed over all SNs (i.e. not dependent on $\x_i$).
If this is not the case (viz. spatially-correlated noise) the ideas provided in~\cite{Ahmadi2018} could be leveraged for removing such constraint.
The \emph{sensing} SNR is defined as $\text{SNR}^\mathrm{s} \triangleq P_t/\sigma^2_s$. Each SN computes its binary local decision, $I(\x_{i})\in\{0,1\}$, by comparing the collected data with a local decision threshold $\tau$, i.e.,
\begin{equation}
\label{eq:Ii}
I(\x_i) = \begin{cases}  1, & g\left( S(\x_i) \right) \geq \tau	\\
                     0, & g\left( S(\x_i) \right) < \tau \end{cases}
\end{equation}
where $g(\cdot)$ is the local detection function, e.g., matched filter or energy detector. 
Here, $\tau$ is assumed to be the same for all SNs for simplicity.
Therefore, the local probabilities of false alarm and detection are given respectively by
\begin{align}
P_{fa}(\x_{i}) & =P_{fa}=f_{0}\left(\tau;\sigma_{s}\right)\label{P_fa}\\
P_{d}(\x_{i}) & =f_{1}\left(\tau;a(\x_{i},\boldsymbol{\theta}),\sigma_{s}\right)\label{eq:P_d}
\end{align}
where $f_0\left(\cdot\,; \sigma_{s}\right)$ and $f_1\left(\cdot\,; a(\x_{i},\boldsymbol{\theta}), \sigma_{s} \right) $ are the complementary cumulative density functions of $g\left( S(\x_i) \right)$ under $\Hyp_0$ and $\Hyp_1$, respectively. 
Both these functions depend on the type of local detector used (matched filter, energy detector, etc.) and the noise level $\sigma_{s}$ (as well as the selected threshold $\tau$).
Additionally, the probability of detection in Eq.~\eqref{eq:P_d} also depends on the target parameters, $\boldsymbol{\theta}$, through Eq.~\eqref{eq:noise-free-signal}.

Due to the large area of the ROI, the WSN is geographically divided into $M$ \emph{disjoint} cluster zones: $\C_1, \C_2, \cdots, \C_M $, where $\C_m \subset \A$ for $m = 1, \cdots, M$.
As a result, each zone $\C_m$ contains a \textit{daughter} PPP, $\Phi_m$, such that
$\Phi = \bigcup_{m=1}^M \Phi_m $.
For simplicity\footnote{We remark that the following results, with some minor modifications, apply even when this simplifying assumption does not hold.}, we approximate the non-homogeneous PPP by choosing the $m$th cluster adequately small so that the intensity within  $\mathcal{C}_m$ is approximately constant, namely $\lambda(\x) \approx \lambda_{m}$ for $\x \in \mathcal{C}_m$, where $\lambda_m$ is the (homogeneous) $m$th cluster SN intensity.
The aforementioned sensing process implies a \emph{thinning} operation for each $\Phi_m$, leading to the (thinned) intensity measure $\lambda_m P_{fa}(x)$ (resp. $\lambda_m P_{d}\,(\x) $) under $\Hyp_0$ (resp. under $\Hyp_1$).
Fig. \ref{fig:WSN_fig} shows a homogeneous random network deployment (i.e. $\lambda_m = \lambda$ for $m=1,\ldots,M$).

\subsection{SN-CH and CH-FC Communication Models}
\label{subsec:com-model}
Each zone is managed by a CH (whose position $\x_m$ does not necessarily fall within $\Phi_m$).
The number of clusters is fixed and their locations are also fixed and known to the WSN. 
\begin{OK}
Accordingly, CH selection is assumed to be preliminary performed based on standard techniques, such as higher computational power/residual energy, minimum distance or combinations of them~\cite{olawole2019fusion}.
Still, we remark that the following analysis applies \emph{independently} on the specific CH selection scheme.
\end{OK}
SNs located at $\x_i \in \Phi_m$ send their decisions to the $m$th CH, over a \emph{shared} channel (due to bandwidth constraints).
The CHs in turn report the collected decisions back over \emph{dedicated} channels to the FC, forming the three-tier network shown in Fig.~\ref{fig:WSN_fig}.
It is assumed that there is an initialization stage in the WSN where synchronization and channel estimation is carried out on the FC and the CHs levels.
Details about both SN-CH and CH-FC links are given in what follows.

\textbf{SN-CH communication:} SNs in the $m$th cluster report to the CH over a (shared) MAC suffering from path-loss with exponent $\alpha$ and a communication reference distance of $r_0$, after which the inverse-power law is valid. 
Depending on the specific value of  $\alpha$, we analyze \emph{two setups} in what follows: ($i$) the free-space propagation model (corresponding to $\alpha = 2$) and ($ii$) the ground-reflection model (corresponding to $\alpha = 4$).

The channel's flat fading gain between the $i$th SN and the  $m$th  CH is $H_{m,i} = \vert H_{m,i} \vert e^{j\varphi_{m,i}}$ where $\vert H_{m,i} \vert$'s are assumed to be i.i.d. Rayleigh random variables (RVs) with parameter $\sigma^2_{H,m}$ and $\varphi_{m,i}$'s are i.i.d. uniform RVs in the interval $[0,2\pi]$. The channels also suffer from AWGN with variance $\sigma^2_{c,m}$.

The SNs estimate the channels with the aid of a (broadcast) pilot signal sent by the CH in the network initialization stage. 
Note, however, that the channels are known to the SNs but not to the CH. This is due to the random number of SNs that makes it difficult to keep track of all the channel gains.

The SNs use on-off-keying (OOK) to send their decisions to the CH over the shared MAC. These SNs transmit with the same power $P_{tx}$ within the cluster and are assumed to be synchronized to the same time slot. The \emph{communication} SNR at $m$th CH is defined as $\text{SNR}^\mathrm{ch}_m \triangleq P_{tx}/\sigma^2_{c,m}$. 

\textbf{CH-FC communication:}
The communication between the CHs and the FC takes place over \emph{dedicated} channels (see Fig.~\ref{fig:WSN_fig}), as it is reasonably assumed that CHs have access to larger bandwidth (than SNs).
Additionally, the fading and path-loss in the CH-FC channels are assumed to be handled (estimated and compensated) in the initialization stage of the network, due to higher available transmit power (denoted with $P_m$).
Hence, it is assumed that the $m$th CH-FC channel only suffers from AWGN with variance $\sigma^2_{f,m}$.
The \emph{communication} SNR between $m$th CH and the FC is then $\text{SNR}^\mathrm{fc}_m \triangleq P_m/\sigma^2_{f,m}$.

%% file: WSN_Fig_Mod.tex
%
%
\newdimen\R
\R=1cm
\newdimen\S
\S=6cm
  
\def\h{1\S}
\def\H{2\S}
\begin{tikzpicture}
%
%
\tikzstyle {SN_style} = [very thick]

\tikzstyle {dSN_style} = [fill = gray, very thick]

\tikzstyle {CH_style} = [fill = gray, very thick, shading=radial]

\tikzset{station/.style={very thick, naming,draw,shape=dart,shape border rotate=90, minimum width=10mm, minimum height=10mm,outer sep=0pt,inner sep=3pt}}

\tikzset{naming/.style={align=center,font=\small}}

\tikzset{antenna/.style={insert path={-- coordinate (ant#1) ++(0,0.25) -- +(135:0.25) + (0,0) -- +(45:0.25)}}}

\draw (0,0) -- (1.8\S,0) -- (2.3\S,.9\S) -- (.5\S,.9\S) -- cycle ++(1,0.2) node{Tier 3}
(5.2,0.3) node[scale=1.4]{$\lambda_1$}
(10.5,0.3) node[scale=1.4]{$\lambda_2$}
(7.5,5) node[scale=1.4]{$\lambda_3$}
(13,5) node[scale=1.4]{$\lambda_4$}
;
%
%
\newcommand{\CHr}[3]{%
\def\L{0.9}
\draw[ultra thick, rounded corners] ++(#1,#2) rectangle  ++(\L,\L) ++(-0.5*\L,-0.5*\L) node(#3){CH} ++(0.5*\L,0);
%
\draw[ultra thick] (#1,#2) ++(\L,0.5*\L)-- ++(4mm,0.0mm) -- +(0mm,4.0mm) -- +(2.625mm,7.5mm) -- +(-2.625mm,7.5mm) -- +(0mm,4.0mm);

\foreach \r in {.3,.6,.9}
      \draw (#1+13mm,#2+12mm) ++ (60:\r) arc (60:120:\r);
}

\newcommand{\CHl}[3]{%
\def\L{0.9}
\draw[ultra thick, rounded corners] ++(#1,#2) rectangle  ++(\L,\L) ++(-0.5*\L,-0.5*\L) node(#3){CH} ++(0.5*\L,0);
%
\draw[ultra thick] (#1,#2) ++(0,0.5*\L)-- ++(-4mm,0.0mm) -- +(0mm,4.0mm) -- +(2.625mm,7.5mm) -- +(-2.625mm,7.5mm) -- +(0mm,4.0mm);

\foreach \r in {.3,.6,.9}
      \draw (#1-4mm,#2+12mm) ++ (60:\r) arc (60:120:\r);
}

\newcommand{\BS}{%
\begin{tikzpicture}[]
\node [station] (base){};

\draw[line join=bevel, very thick, scale = 1.5] (base.110) -- (base.70) -- (base.north west) -- (base.north east) -- cycle;
\draw[line join=bevel, very thick, scale = 1.5] (base.100) -- (base.80) -- (base.110) -- (base.70) -- (base.north west) -- (base.north east);
\draw[line join=bevel, very thick, scale = 1.5] (base.100) -- (base.70) (base.110) -- (base.north east);

\draw[line cap=rect,, very thick, scale = 1.5] ([yshift=0pt]base.north) [antenna=1];

\end{tikzpicture}
}
%
%

\draw[dashed] (0.25\S,0.45\S) -- (2.05\S,0.45\S);
\draw[dashed] (0.9\S,0.0\S) -- (1.35\S,0.9\S);

\foreach \xcen/\ycen in { 1.8/0.5, 2.5/2, 3/1, 4.1/1.8, 4.7/0.6}
		\draw[SN_style] (\xcen, \ycen) circle (0.2cm);

\foreach \xcen/\ycen in {1.2/1.4, 4.8/2.4, 5.6/1.3}{
		\draw[dSN_style] (\xcen, \ycen) circle (0.2cm);
		\draw[->, very thick, dashed] (\xcen, \ycen) -- (0.62\S,0.225\S+\h);
		}

\foreach \xcen/\ycen in {2.3/3, 3/4, 4.1/3.1, 5.4/4.3, 5.1/3.3, 6.1/3.8, 3.7/4.8}
		\draw[SN_style] (\xcen, \ycen) circle (0.2cm);

\draw (3.2,4) node[right]{$\text{SN}$};

\foreach \xcen/\ycen in {6.1/3.8, 6.5/3.0}
		\draw[dSN_style] (\xcen, \ycen) circle (0.2cm);		

\foreach \xcen/\ycen in {8.3/4.8, 9.7/4.6, 10.1/3.1, 10.7/5.1, 11.6/4.1, 12.2/5.1}
		\draw[SN_style] (\xcen, \ycen) circle (0.2cm);
		
\foreach \xcen/\ycen in {7.7/3.9, 8.5/3.3, 11.2/3.0}{
		\draw[dSN_style] (\xcen, \ycen) circle (0.2cm);
		}

\foreach \xcen/\ycen in {6.8/0.5, 8.2/1.3, 8.1/2.1, 9.2/2.0, 9.4/0.8, 10.1/1.6, 11.2/2.2}{
		\draw[SN_style] (\xcen, \ycen) circle (0.2cm);
		}
		
\foreach \xcen/\ycen in {7.2/1.5}
		\draw[dSN_style] (\xcen, \ycen) circle (0.2cm);
				
\tikzstyle{target}=[star, star points=5, star point ratio=2.25, draw,inner sep=0.15em,anchor=outer point 3, fill = red]
\draw[very thick] (5.7,2) node[target]{} ;
		
\draw (0,0+\h) -- (1.8\S,0+\h) -- (2.3\S,.9\S+\h) -- (.5\S,.9\S+\h) -- cycle (1,1.03\h) node{Tier 2};

\draw[dashed] (0.25\S,0.45\S+\h) -- (2.05\S,0.45\S+\h);
\draw[dashed] (0.9\S,0.0\S+\h) -- (1.35\S,0.9\S+\h);

\CHr{0.55\S}{0.225\S+\h}{CHa}

\CHl{0.55\S+0.9\h}{0.225\S+\h}{CHb}

\CHl{0.5\S+1.1\h}{0.225\S+1.4\h}{CHc}
\draw (0.72\S+0.5\h, 0.35\S+0.8\h)+(-1.2,0) node[right,scale=1.5]{$Y_m=\sqrt{P_{tx}}\bar{Y}_m+W_m$};

\CHr{0.68\S}{0.225\S+1.4\h}{CHd}

\draw (0,0+\H) -- (1.8\S,0+\H) -- (2.3\S,.9\S+\H) -- (.5\S,.9\S+\H) -- cycle ++(1,0.2) node{Tier 1};

\draw (1.15\S,0.6\S+\H) node[scale=1.8](BS){\BS};
\draw (1.15\S, 0.6\S+\H -11mm) node[above]{$\text{FC}$};
\draw (1.25\S, 0.5\S+\H) node[right,scale=1.5]{$Z_m = \sqrt{P_m}Y_m+V_m$};

\foreach \ch in {CHa, CHb, CHc, CHd}{
    \draw[->, very thick, dashed] (\ch.north)  -- (BS);
}

\end{tikzpicture}

%% file: 5_dist_trans_comb_rev2.tex
\section{Distributed Transmit Combining for Multiple Access Clustered Distributed Detection}
\label{sec:Dist-Trans-Comb}

In this section, first the distributed transmit combining techniques employed in this work are discussed (Sec.~\ref{subsec:dist-Tx-comb}).
Then, the associated statistics of the received signal are obtained (Sec.~\ref{subsec:received-signal-pdf}). 
The latter result paves the way to the formulation of the fusion rules in the next section. 

\subsection{Distributed Transmit Combining}
\label{subsec:dist-Tx-comb}
Although adopting MACs at the cluster level significantly reduces bandwidth requirements, the received signals at the CHs do not benefit from the beamforming-like feature of the MAC due to fading~\cite{Liu2007a}. Clearly, conventional receive combining techniques cannot be implemented in the MAC case. However, \emph{transmit combining schemes} can be used. In fact, such schemes can be realized in a distributed manner by virtue of the shared MAC, since all the transmitted signals are combined at each CH.

Accordingly, in this work we consider \emph{two} distributed transmit combining methods: ($i$) the distributed maximum ratio transmit combining (dMRTC) and ($ii$) the distributed equal gain transmit combining (dEGTC). 
The dMRTC is implemented if the SNs pre-multiply their transmitted signals by the complex channel gain i.e., $G_{m,i} = H^*_{m,i}$, where the channel gain and phase are estimated in the initialization stage. Whereas the dEGTC is implemented if the SNs adjust the transmitted signal phase, i.e.,  $G_{m,i} = e^{-j\varphi_{m,i}}$, where only the channel phase is estimated.
\begin{OK}
 Clearly, the dMRTC implementation requires both channel gain and phase estimation leading to more complexity in the system when compared to the dEGTC, which requires only the phase estimation.
\end{OK}
In order to represent both cases in a compact fashion, we define the following generic mapping:
\begin{equation}
f(H_{m,i})=H_{m,i}\,G_{m,i}=\begin{cases}
\vert H_{m,i}\vert^{2}, & \mbox{dMRTC}\\
\vert H_{m,i}\vert, & \mbox{dEGTC}
\end{cases}.\label{eq:dTC}
\end{equation}

As a result, the received signal at the $m$th CH is
\begin{equation}
Y_{m}=\sqrt{P_{tx}}\,\underbrace{\sumCm\frac{f(H_{m,i})}{\Xixm}I\left(\Xbi\right)}_{\triangleq\bar{Y}_{m}}\,\,+\,\,W_{m}\label{eq:Y}
\end{equation}
for $ m = 1,\cdots,M $, $W_m$ is the AWGN at that CH with distribution of $\N\left(0,\sigma^2_{c,m} \right)$ and $\bar{Y}_m$ denotes the power-scaled noise-free contribution, defined to simplify the analysis later on.

The received signals at the FC from all the $M$ CHs are 
\begin{equation}
Z_m = \sqrt{P_m} \, Y_m + V_m, \; m = 1,\cdots,M \label{eq:CH-FC}
\end{equation}
where $P_m$ is the transmission power used by the $m$th CH and $V_m \sim \N\left(0,\sigma^2_{f,m} \right)$ is the AWGN associated to the channel between the $m$th CH and the FC.
 In order to develop the optimal fusion rule in clustered WSNs with noisy channels, we investigate the received signals at the FC. 
By combining Eqs.~\eqref{eq:Y}~and~\eqref{eq:CH-FC}, the received signal from $m$th CH can be rewritten in the more convenient form as
\begin{equation}
Z_m = \sqrt{\Pt_m} \bar{Y}_m + \Vt_m\;, m = 1,\cdots,M \label{eq:Z_m}
\end{equation}
where $\Pt_m \triangleq P_{tx} \, P_m$, and $ \Vt_m \triangleq \sqrt{P_m} W_m + V_m$ denotes the aggregate transmission power and noise at the $m$th CH-FC channel respectively, with distribution $\N\left( 0, \sigmat^2_m \right)$ where $\sigmat^2_m \triangleq (P_m \, \sigma^2_{c,m} + \sigma^2_{f,m})$.

\subsection{Received (Noise-free) Signals Statistics}\label{subsec:received-signal-pdf}
The noiseless received signal $\bar{Y}_m$ in Eq. \eqref{eq:Y} is actually a random sum over the point process of detecting SNs. Unfortunately, its distribution does not have a closed-form. Nonetheless, the mean and variance of $\bar{Y}_m$ can be found via stochastic-geometry tools.
Firstly, the mean is given below as

\begin{align}\label{eq:mean-sum}
\mubmj & =\E\left[\bar{Y}_{m}|\H_{j}\right]=\E\left[\sumCm\frac{f\left(H_{m,i}\right)}{\Xixm}\I\bigg\rvert\H_{j}\right]\nonumber \\
 & =\E\left[f\left(H\right)\right]\E_{\Phi_{m}}\left[\sumCm\frac{\I}{\Xixm}\bigg\rvert\H_{j}\right]
\end{align}
where $\E_{\Phi_m}\left[\cdot\right]$ is the expectation with respect to PPP $\Phi_m$ and  $j=0$ (resp. $j=1$) denotes the $\H_{0}$ (resp. $\H_{1}$) hypothesis. 
The conditional mean $\mubmj$ can be further simplified as demonstrated by the following proposition.

\begin{prop} \label{PROP:MEAN}
The conditional mean of $\bar{Y}_m$ defined in Eq. \eqref{eq:Y} is given by
\begin{equation}
\mubmj=\E\left[\bar{Y}_{m}|\H_{j}\right]=\begin{cases}
\lambda_{m}\,\,\E\left[f\left(H\right)\right]\Imu{0}, & j=0\\
\lambda_{m}\,\,\E\left[f\left(H\right)\right]\Imu{1}, & j=1
\end{cases}\label{eq:mean-int}
\end{equation}
where 
\begin{eqnarray}
\Imu{0} & \triangleq &  \int\limits_{\C_m}\Vert \x - \x_m \Vert^{-\frac{\alpha}{2}} \, P_{fa} \, d\xb \label{eq:I_mu_0}  \\
\Imu{1} & \triangleq &  \int\limits_{\C_m}\Vert \x - \x_m \Vert^{-\frac{\alpha}{2}} \, \Pdx \, d\xb .\label{eq:I_mu_1}
\end{eqnarray}
\end{prop}
\begin{IEEEproof}
Recalling that the local detection is actually a thinning of the PPP, then Campbell's theorem~\cite{Streit2010} can be applied to find the average of the expectation in Eq.~\eqref{eq:mean-sum} yielding the result in Eq.~\eqref{eq:mean-int}.
\end{IEEEproof}
The computation of the conditional variance, on the other hand, is not as straightforward. The following proposition provides its explicit value.
\begin{prop} \label{PROP:VAR}
The conditional variance of $\bar{Y}_m$ defined in Eq. \eqref{eq:Y} is given by
\begin{gather}
\sigbmj=\mathrm{var}\left(\bar{Y}_{m}|\H_{j}\right)=\begin{cases}
\lambda_{m}\,\E\left[f^{2}(H)\right]\Isig{0}, & j=0\\
\lambda_{m}\,\E\left[f^{2}(H)\right]\Isig{1}, & j=1
\end{cases}\label{eq:var_j}
\end{gather}
where
\begin{eqnarray}
\Isig{0} & \triangleq & \displaystyle \int\limits_{\C_m} \Vert \x - \x_m \Vert^{-\alpha} \, P_{fa} \, d\xb \label{eq:I_sigma2_0} \\
\Isig{1} & \triangleq & \displaystyle \int\limits_{\C_m} \Vert \x - \x_m \Vert^{-\alpha} \, \Pdx \, d\xb \label{eq:I_sigma2_1}
\end{eqnarray}
and $\E\left[f^{2}(H)\right]$ denotes the second (non-central) moment of $f(H)$.
\end{prop}
\begin{IEEEproof}
See Appendix \ref{sec:App-A}.
\end{IEEEproof}

%% file: 6_dist_det_multiple_clusters_rev3.tex
\section{Fusion Rules for Distributed Detection in Multiple Clusters}\label{sec:Multiple-Clusters}
In this section we derive four fusion rules for distributed detection in multiple clusters based on approximating the received signal distribution by Gaussian and lognormal distributions. 
Note however, that both fusion rules can be used with either transmit combining techniques described in Eq.~\eqref{eq:dTC}.

\subsection{Optimal Fusion Rule (LLR)}
The Neyman-Pearson detector \cite{Kay1998}, which is based on the likelihood-ratio-test (LR) statistic, for the model in Eq. \eqref{eq:Z_m} is 
\begin{gather}
\Lambda_{\text{LR}}=\prod_{m=1}^{M}\frac{p\left(z_{m}|\Hyp_{1}\right)}{p\left(z_{m}|\Hyp_{0}\right)}=\prod_{m=1}^{M}\frac{\E_{\bar{Y}_{m}|\mathcal{H}_{1}}\left[p\left(z_{m}\vert \bar{Y}_{m}\right)\right]}{\E_{\bar{Y}_{m}|\mathcal{H}_{0}}\left[p\left(z_{m}\vert \bar{Y}_{m}\right)\right]}\nonumber \\
=\prod_{m=1}^{M}\frac{\E_{\bar{Y}_{m}|\mathcal{H}_{1}}\left[\exp\left(-\frac{1}{2\sigmat_{m}^{2}}(z_{m}-\sqrt{\Pt_{m}}\bar{Y}_{m})^{2}\right)\right]}{\E_{\bar{Y}_{m}|\mathcal{H}_{0}}\left[\exp\left(-\frac{1}{2\sigmat_{m}^{2}}(z_{m}-\sqrt{\Pt_{m}}\bar{Y}_{m})^{2}\right)\right]}.\label{eq:Noisy-LRT}
\end{gather}

Note that the expectations in the numerator and denominator are w.r.t. the distributions $p(\bar{y}_{m}|\mathcal{H}_{j})$, resulting from PPP thinning and nonlinear mapping.
Indeed,  $p\left( z_m | \Hyp_j \right)$ is actually the convolution of the distribution of $\sqrt{\Pt_{m}} \cdot \bar{Y}_m |  \Hyp_j $ and the Gaussian distribution of the noise $\Tilde{V}_m$. 
Unfortunately, the corresponding log-likelihood ratio (LLR) $\LLLR  \triangleq  \ln\left( \Lambda_{\text{LR}} \right)$ is not simpler:
\begin{eqnarray}
\LLLR & = & \sum_{m=1}^{M}\ln\left(\E_{\bar{Y}_m|\mathcal{H}_{1}}\left[\exp\left(-\frac{s_{m}}{2}\left(\zt_{m}-\bar{Y}_m\right)^{2}\right)\right]\right)\nonumber \\
 & - & \ln\left(\E_{\bar{Y}_m|\mathcal{H}_{0}}\left[\exp\left(-\frac{s_{m}}{2}\left(\zt_{m}-\bar{Y}_m\right)^{2}\right)\right]\right)\label{eq:LLR-Noisy}
\end{eqnarray}
where $\zt_m = z_m / \sqrt{\Pt_m}$ and $s_m = \Pt_m / \sigmat^2_m$ is the $m$th CH-FC equivalent channel SNR.

\subsection{Moment Matching based Fusion Rules}

Although the fusion rule in Eq. \eqref{eq:LLR-Noisy} is optimal, unfortunately it is impractical and does not lend itself to analysis.
Accordingly, to come up with the design of practical fusion rules, we provide a second-order characterization of the received signals $Z_{m}|\mathcal{H}_{j}$, $m=1,\ldots,M$.

Based on the results provided in Props.~\ref{PROP:MEAN}~and~\ref{PROP:VAR}, it is not difficult to show that the mean $\left(\mu_{m,j}\triangleq \mathbb{E}\left[Z_{m}|\mathcal{H}_{j}\right] \right)$ can be obtained (by linearity) as
\begin{align}
\mu_{m,j}  \triangleq \mathbb{E}\left[Z_{m}|\mathcal{H}_{j}\right] 
& =\sqrt{\Pt_{m}}\,\mathbb{E}\left[\bar{Y}_{m}\,|\,\mathcal{H}_{j}\right]+\mathbb{E}\left[\Vt_{m}\right] \nonumber \\
& =\sqrt{\Pt_{m}}\,\mubmj.
\end{align}

Conversely, the variance evaluation follows as
\begin{align}
\sigma_{m,j}^{2}  \triangleq \mathrm{var}\left(Z_{m}\,|\,\mathcal{H}_{j}\right) & = \Pt_{m}\,\mathrm{var}\left[\bar{Y}_{m}\,|\,\mathcal{H}_{j}\right]+\mathrm{var}\left[\Vt_{m}\right] \nonumber \\
& = \Pt_{m}\,\sigbmj+ P_m \sigma^2_{c,m} + \sigma^2_{f,m}
\end{align}
which is the aggregate variance of the CHs' received signals, the SN-CH links and the CH-FC links as well.
Having found the mean and variance of $Z_m|\mathcal{H}_j$, it is possible to approximate its (conditional) distribution, for any $m=1,\ldots,M$, via the moment matching method.

The lognormal distribution was adopted in \cite{Aldalahmeh2019a} for fitting due to having two defining parameters and hence it can fit the $Y_m$'s distribution, which is suitable for high SNR cases since the lognormal distribution is defined on a positive support. In this work we relax this condition on the SNR and choose the Gaussian distribution for moment matching. We will derive the the moment matching fusion rule for the Gaussian case in Theorem \ref{THM:MOFR-N} and then for the lognormal in Theorem \ref{THM:MOR-L}.

Firstly, using the approximated distributions (Gaussian or lognormal) the LLR can be expressed without the expectations in Eq. \eqref{eq:LLR-Noisy}. Accordingly, the following theorem provides the moment matching optimal fusion rule (MOR) when adopting the Gaussian distribution fitting,
%
%
\begin{thm}
\label{THM:MOFR-N}
The MOR detector, in the Neyman-Pearson sense, using the Gaussian distribution fitting
is given by
\begin{equation}
\Lambda^N_{\mathrm{MOR}} = \sum_{m=1}^M a_m \left( z_m + d_m \right)^2
\label{eq:NP-FR}
\end{equation}
where
\begin{eqnarray}
a_m & \triangleq & \frac{1}{2\sigma^2_{m,0}} - \frac{1}{2\sigma^2_{m,1}} \label{eq:a_m}\\
d_m & \triangleq &  \frac{ \sigma^2_{m,0} \mu_{m,1} - \sigma^2_{m,1} \mu_{m,j}}{\sigma^2_{m,j} - \sigma^2_{m,j} }\label{eq:c_m}
\label{eq:mu_m}\; 
\end{eqnarray}
for $j = 0,1$. 
\end{thm}
\begin{IEEEproof}
See Appendix \ref{sec:App-MOR-N-Proof}.
\end{IEEEproof}

The optimal detector given above can be regarded as the Euclidean distance between the received CHs data, $Z_m$'s, and the points $d_m$'s, which is \textit{weighted} by $a_m$'s.

If the communication SNR is suitably high, we conjecture that the lognormal distribution might be adequate for developing a fusion rule. This is presented in the following theorem.
%
%
\begin{thm}
\label{THM:MOR-L}
The MOR detector using the lognormal distribution fitting is given by
\begin{equation}
\Lambda^L_{\mathrm{MOR}} = \sum_{m=1}^M \hat{a}_m \left( \ln \vert z_m \vert + \hat{d}_m \right)^2
\label{eq:MOF-L}
\end{equation}
where
\begin{eqnarray}
\hat{a}_m & \triangleq & \frac{1}{2\sighmj{0}} - \frac{1}{2\sighmj{1}} \label{eq:a_m}\\
\hat{d}_m & \triangleq &  \frac{ \sighmj{1} \muhmj{0} - \sighmj{0} \muhmj{1}}{\sighmj{1}- \sighmj{0} }\label{eq:c_m} 
\end{eqnarray}
for $j = 0,1$.

\end{thm}
\begin{IEEEproof}
See Appendix \ref{sec:App--MOR-L-Proof}.
\end{IEEEproof}
Note that the structure of the fusion rule in Eq. \eqref{eq:MOF-L} is similar to that derived in \eqref{eq:MER-N}. This is due to the fundamental similarity between the Gaussian and lognormal distributions. 

\subsection{Moment Matching Equal Gain Fusion Rule}
Clearly, the optimal fusion rules above require knowledge of the CH's received signal statistics to compute the parameters $a_m$ and $d_m$, which in turn require the knowledge of the target's parameters, a case that is not always available in practice. Consequently, we propose using the same moment matching-based statistics but with equally weighing all the clusters data, termed moment matching equal gain fusion rule (MER) as
\begin{equation}
\Lambda^L_{\text{MER}} = \SumM \ln^2 \vert z_m \vert.
\label{eq:MER-L}
\end{equation}

The MER simply takes the logarithm of each CH transmitted signal and sums them together. 
 Similarly, the MER for the Gaussian fitting case is given as
\begin{equation}
\Lambda^N_{\text{MER}} = \SumM z_m^2
\label{eq:MER-N}
\end{equation} 
which is the sum of the squares of the received signals from the CHs.

%% file: 7_Performance_analysis_rev3.tex
\section{Performance and Power Analysis}
\label{sec:Perf-analys}
Analyzing the performance of the multiple clusters fusion rule is not straightforward, so we defer this work to future works due to space constraints. However, we consider the effect of the distributed combining techniques on the the detection performance of the single cluster case and conjecture that the same applies to the multiple clusters case. Fortunately, this conjecture is validate by the simulation results in Sec.\ref{sec:Simulation-Result}. In contrast, the effect of clustering on the transmitted and received power is analyzed. 

\subsection{Single Cluster Distributed Detection}
In the single cluster case, the system is modeled by Eq. \eqref{eq:Y}. For the sake of consistency, let us denote the single cluster case with the  zero cluster index, i.e., $m = 0$. The received signal now is $Y_0$.
The latter FC reaches its global decision on the target's presence by comparing the received signal with a global detection threshold, $\Gamma$.
For the general case when the received signal is approximated by a Gaussian distribution.
The global detection performance can be readily found. The global probability of false alarm is
\begin{equation}
P_{FA}  =  \mathbb{P}\left(Y>\Gamma;\H_{0}\right) = \mathcal{Q}\left(\frac{\Gamma-\mu_{0,0}}{\sigma_{0,0}}\right)
\end{equation}
where $\mathcal{Q}(\cdot)$ is the error Q-function, $\mu_{0,0}$ and $\sigma_{0,0}$ are the mean and standard deviation of $Y_0$ under $\Hyp_0$ as defined in Eq. \eqref{eq:mean-int}. Note that the single cluster encompasses the ROI, i.e.,  $\mathcal{C}_0 = \mathcal{A}$.
Consequently, given $P_{FA}$ the global detection threshold can be found as $\Gamma=\sigma_{0,0}\mathcal{Q}^{-1}\left(P_{FA}\right)+\mu_{0,0}$. 

On the other hand, the global probability of detection is
\begin{eqnarray}
P_{D} & = & \mathbb{P}\left(Z>\Gamma;\H_{1}\right) = \mathcal{Q}\left(\frac{\Gamma-\mu_{0,1}}{\sigma_{0,1}}\right)\nonumber \\
 & = & \mathcal{Q}\left(\frac{\mu_{0,0}-\mu_{0,1}+\sigma_{0,0}\mathcal{Q}^{-1}\left(P_{FA}\right)}{\sigma_{0,1}}\right) \label{eq:Q-func}
\end{eqnarray}
where $\mu_{0,1}$ is the mean of $Y_0$ under $\Hyp_1$.
Unfortunately, Eq. \eqref{eq:Q-func} does not provide an insight into the performance of the detector due to the complications in Eqs. \eqref{eq:mu_a} and  \eqref{eq:sigma2_a} w.r.t. $\lambda$ and $P_{tx}$. Therefore, we choose to investigate the \emph{deflection coefficient}~\cite{Kay1998} in terms of the means and variances given by Props.~\ref{PROP:MEAN}~and~\ref{PROP:VAR}, which is
\begin{equation}
d^2 =  \frac{\left( \mu_{0,1} - \mu_{0,0} \right)^2}{\sigma^2_{0,1}}. \label{eq:d2}
\end{equation}

Substituting Eqs. \eqref{eq:mean-int} and \eqref{eq:var_j} in Eq. \eqref{eq:d2} yields
\begin{equation}
d^{2}=\lambda g_{tc}\frac{\left(I_{\bar{\mu}_{0,1}}-I_{\bar{\mu}_{0,0}}\right)^{2}}{I_{\bar{\sigma}_{0,1}^{2}}}\label{eq:d2-fin}
\end{equation}
where 
\begin{equation}
g_{tc} = \frac{\E^2\left[ f(H) \right] }{\E\left[ f^2(H) \right] }.
\end{equation}

Note that the deflection coefficient above depends on network deployment through $\lambda$ and the local detector through 
$I_{\bar{\mu}_{0,0}}, I_{\bar{\mu}_{0,1}}$ and $I_{\bar{\sigma}_{0,1}^{2}}$.
Also, the effect of the transmit combining scheme appears in $g_{tc}$, which we term as the transmit combining gain. For the dMRTC case, the gain is 
\begin{equation}
g_{tc} = \frac{4\sigma^4_H}{4\sigma^4_H+4\sigma^4_H} =\frac{1}{2}
\end{equation}
where the second moment in the denominator is given by the identity $\E[X^2] = \sigma^2_X + \E^2[X]$ and the mean and variance are that of the exponential distribution. In a similar manner, the gain under the dEGTC case is 
\begin{equation}
g_{tc} =  \frac{0.5\pi\sigma^2_H}{(2-0.5\pi)\sigma^2_H + 0.5\pi\sigma^2_H} = \pi/4
\end{equation}
where the distribution at hand is Rayleigh. Thus
\begin{equation}
g_{tc}=\begin{cases}
1/2, & \mbox{dMRTC}\\
\pi/4, & \mbox{dEGTC}
\end{cases}
\label{eq:Deflection-gain}
\end{equation}
implying that the dEGTC has a better gain when compared to dMRTC, which is contrary to the well-known case of receiver combining in wireless communication, since the deflection coefficient in the dEGTC is larger. This result can be explained in the context of distributed detection as having better separation between the received signal distributions under $\Hyp_0$ and $\Hyp_1$ in the dEGTC case compared to the dMRTC as predicted by the deflection coefficient and the gain defined in Eq. \eqref{eq:Deflection-gain}.

\subsection{Received Power Analysis}
In PAC WSNs, in which every SN has a dedicated channel, it is known that the SN transmission power must be increased when the cluster size is large in order to keep the SNR at the CH constant. However, this is not case in multiple access channels. In fact it is the exact opposite, as we shall show now.
To this end, we consider the received power under $\Hyp_0$, since the detection event is rare. 
The average received power then is given by the following proposition. 
\begin{prop}
\label{prop:Prx_ave}
The average received power at the $m$th CH under $\Hyp_0$ is 
\begin{equation}
	\bar{P}_{rx,m} = \lambda P_{tx} \left( \E \left[ f^2(H)\right] I^2_{\bar{\sigma}^2_{m,0}} + \lambda \E^2\left[ f(H)\right] \Imu{0} \right).
\label{eq:Prx_ave}
\end{equation}

\end{prop}
\begin{proof}
Recall that the received power at the $m$th CH is $P_{tx} \bar{Y}^2_m$, which is a RV due to fading and random SNs deployment. The average value is 
\begin{equation}
\bar{P}_{rx,m}=P_{tx}\,\E\left[\bar{Y}_{m}^{2}\rvert\Hyp_{0}\right]=P_{tx}\,\bar{\mu}_{m,0}^{2}+P_{tx}\,\bar{\sigma}_{m,0}^{2}.\label{eq:P_rx_ave}
\end{equation}
The proof is concluded when the Eqs.~\eqref{eq:mean-int}~and~\eqref{eq:var_j} are substituted in Eq. \eqref{eq:P_rx_ave}.
\end{proof}
In order to gain insight into the received CH power, analytical forms of $\bar{\mu}_{m,0}$ and $\bar{\sigma}_{m,0}^{2}$  are required, 
which involves the solution of the integrals in Eqs. \eqref{eq:I_mu_0}, \eqref{eq:I_mu_1}, \eqref{eq:I_sigma2_0} and \eqref{eq:I_sigma2_1}.
Unfortunately, this might not be straightforward provided the square cluster shape. Hence, we find the above integrals for circular clusters that encompasses the actual cluster, where the radius equals the distance from the CH to the square cluster corner.
The following corollary provides an approximate close-form power expression.
\begin{corollary}
\label{cor:Prx_circ}
The average received power for a disk-shaped cluster with outer radius of $R$ and inner radius of $r_0$ (the communication reference distance) is given by 
\begin{align}
\bar{P}^c_{rx} & \approx \begin{cases}
 P_{tx} \left( \lambda  K_1  \ln R  + \lambda^2  K_2   R^2 \right), & \alpha = 2 \\
P_{tx} \left(  \dfrac{3 \lambda K_1}{r^2_0}  + \lambda^2 K_2  \ln^2 R  \right), & \alpha = 4 
\label{eq:Prx_circ_app}
\end{cases}
\end{align}	
where the constants are
\begin{align}
	K_1 & = 2\pi P_{fa} \E \left[ f^2(H)\right] \label{eq:K1} \\
	K_2 & = 4\pi^2 P^2_{fa} \E^2 \left[ f(H)\right].\label{eq:K2}
\end{align}

Furthermore, the received power scales as
\begin{equation}
\bar{P}^c_{rx} \sim \begin{cases} 
						\mathcal{O}\left( \lambda^2 R^2 \right)  , & \alpha = 2 \\
						 \mathcal{O}\left( \lambda^2 \ln^2 R \right)  , & \alpha = 4
					\end{cases}.
\label{eq:Prx_inv_R}					
\end{equation}

\end{corollary}
\begin{proof}
    We first note that under $\Hyp_0$ we have $\bar{\mu}_{m,0}$'s are equal for all $m$ and so are $\bar{\sigma}_{m,0}^{2}$'s, hence any cluster will suffice in solving the integrals in Eq. \eqref{eq:Prx_ave}. Given circular clusters, employing polar coordinates yields 
    \begin{align}
& \bar{P}^c_{rx}  = \!\! \begin{cases}
\lambda P_{tx} \left(   K_1  \ln \left( \dfrac{R}{r_0} \right) + \lambda  K_2  \left( R - r_0 \right)^2 \right), &  \alpha = 2 \\
\lambda P_{tx} \left( 3  K_1 \left( \dfrac{R^2 - r^2_0}{R^2 r^2_0} \right) + \lambda K_2  \ln^2 \left( \dfrac{R}{r_0} \right)  \right), & \alpha = 4
\end{cases} \label{eq:Prx_circ_exact}
\end{align}
where $K_1$ and $K_2$ are defined in Eqs. \eqref{eq:K1} and \eqref{eq:K2}. If $R \gg r_0$ then Eq. \eqref{eq:Prx_circ_app} follows directly from Eq. \eqref{eq:Prx_circ_exact}.
\end{proof}

It is evident from the corollary that the received power increases when the cluster size increases, which is the opposite to the PAC case in both free-space path-loss and ground-reflection cases.
This is explained firstly by having more transmitting SNs in the CH as it expands. Secondly, the use of distributed transmit combining techniques (dEGTRC and dMRTC). Finally, the aggregation of the received signals at the CHs due to the MAC nature. Those factors overcome the negative effects of path-loss and fading in the channel.

Another direct result from Eq. \eqref{eq:Prx_inv_R} is that the received power is inversely proportional to the number of clusters. This follows when having uniform clustering with $M$ clusters in a square ROI with side length of $A$, then the circular cluster radius is $R = A/M$ and consequently the received power reduces as $M$ increases.

From the system design point of view, it is desirable to have a specific SNR at the CH receiver. Thus, the transmission power should be increased when having more clusters. The CH's received SNR can be approximated as $\text{SNR}^\mathrm{ch} \triangleq \bar{P}^c_{rx} / \sigma^2_c$, where $\sigma^2_{m,c} = \sigma^2_c\, \forall m$ under $\Hyp_0$. As a result, the SN transmission power can be approximated as
\begin{align}
& P_{tx} \approx  \begin{cases}
  \dfrac{\sigma^2_c \text{SNR}^\mathrm{ch}}{\lambda  K_1  \ln R + \lambda^2  K_2  R^2} , & \alpha = 2 \\
 \dfrac{\sigma^2_c \text{SNR}^\mathrm{ch} }{ \dfrac{3 \lambda K_1}{ r^2_0}  + \lambda^2 K_2  \ln^2 R}  , & \alpha = 4 
\end{cases}
\label{eq:Ptx_app}
\end{align}	
where it is clear now that for a given fixed SNR at the CH, the transmission power is inversely proportional to distance to the CH. Interestingly however, the power is also inversely proportional to the deployment density. So a higher density leads to lower SN transmission power for a fixed SNR.
%

%% file: 8_simulation_results_rev3.tex
%
%
%
%
\begin{figure*}[!t]
\def\figTwoScl{0.18}
\def\hgt{0.19\textwidth}
\centering 
\subfloat[Received signal mean at $\alpha = 2$.]{
\includegraphics[width=0.45\textwidth,height=\hgt]{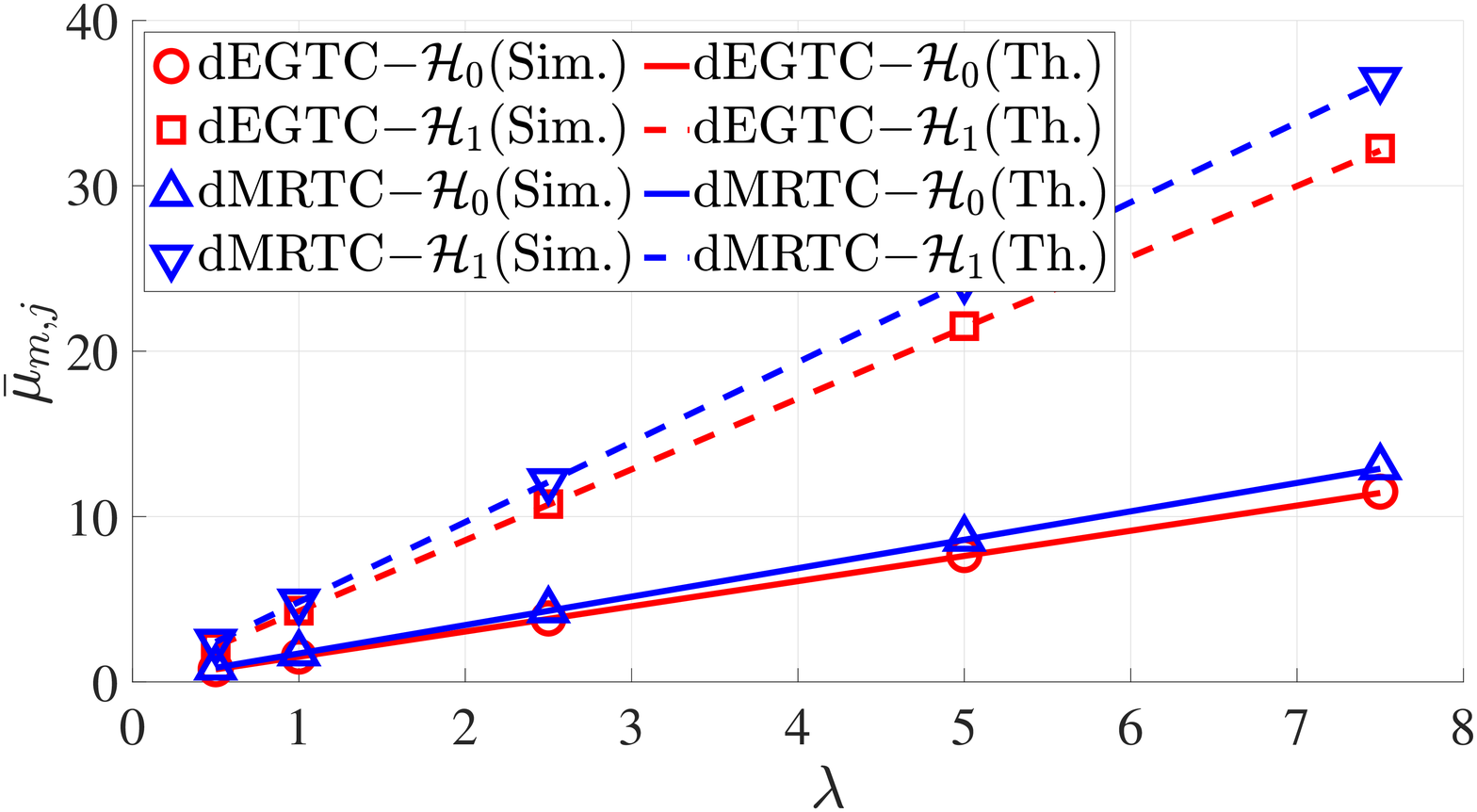}}
\hfill
\centering
\subfloat[Received signal variance at $\alpha = 2$.]{ 
\includegraphics[width=0.45\textwidth,height=\hgt]{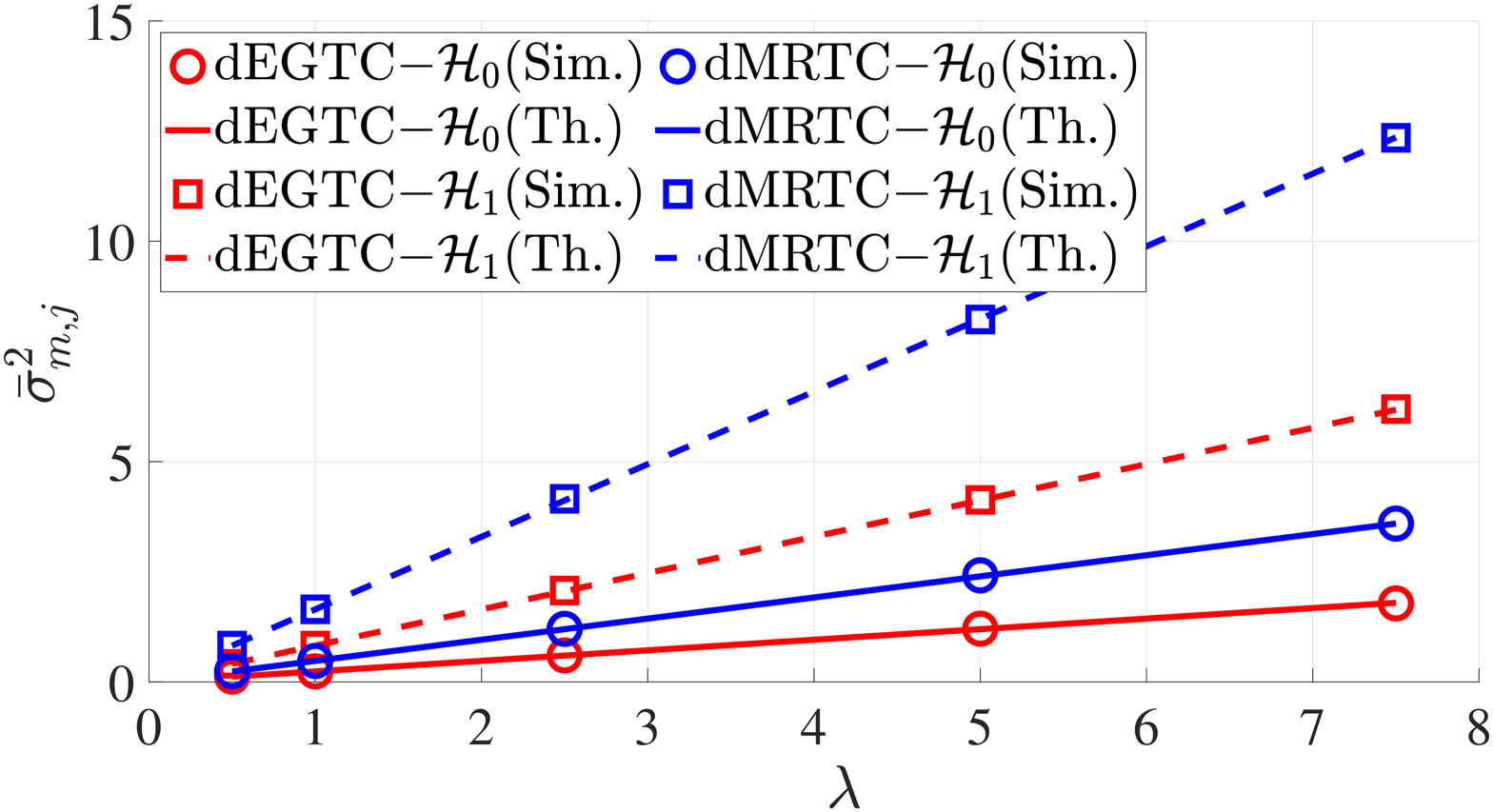}}
\hfill
\centering
\subfloat[Received signal mean at $\alpha = 4$.]{ 
\includegraphics[width=0.45\textwidth,height=\hgt]{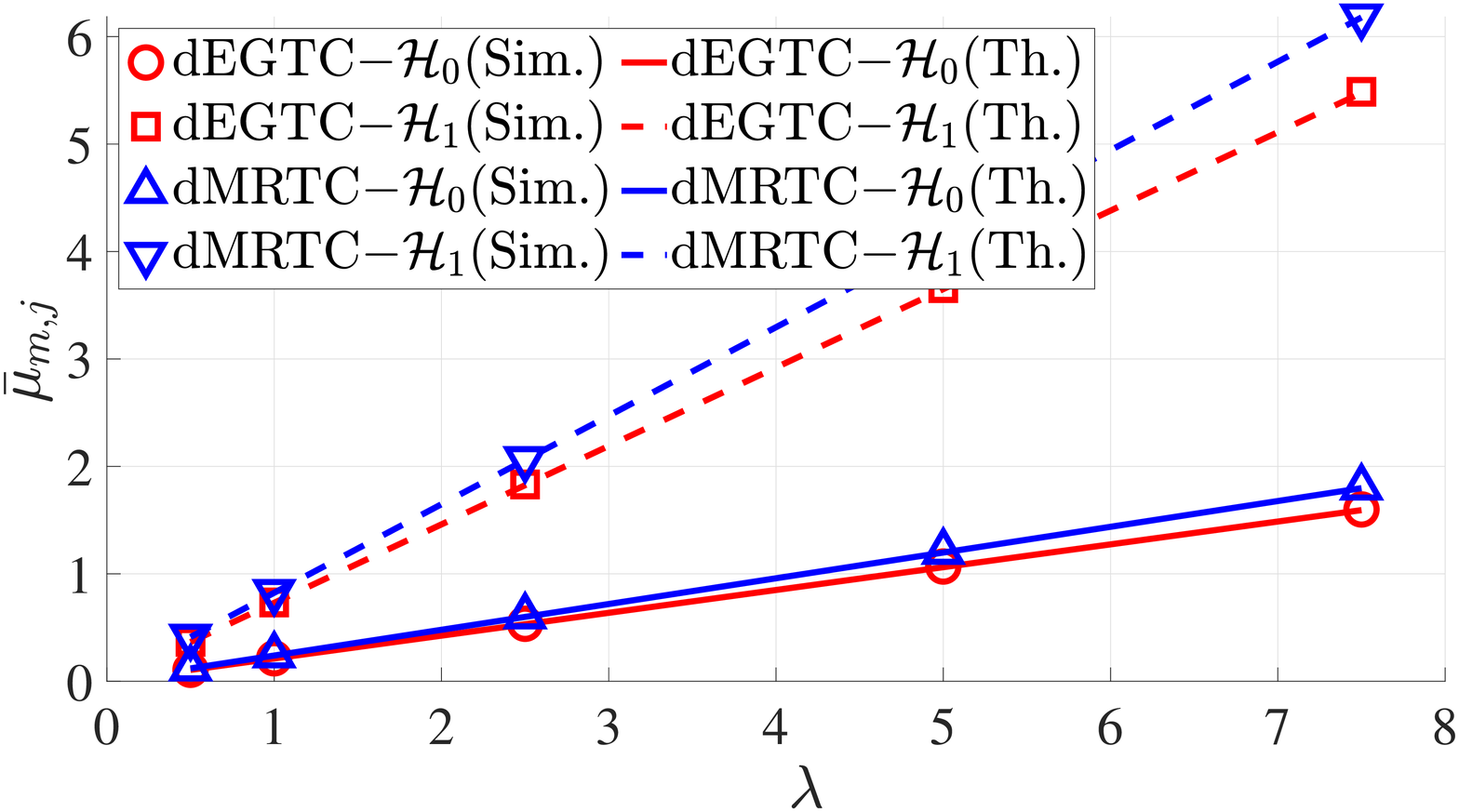}}
\hfill
\centering
\subfloat[Received signal variance at $\alpha = 4$.]{ 
\includegraphics[width=0.45\textwidth,height=\hgt]{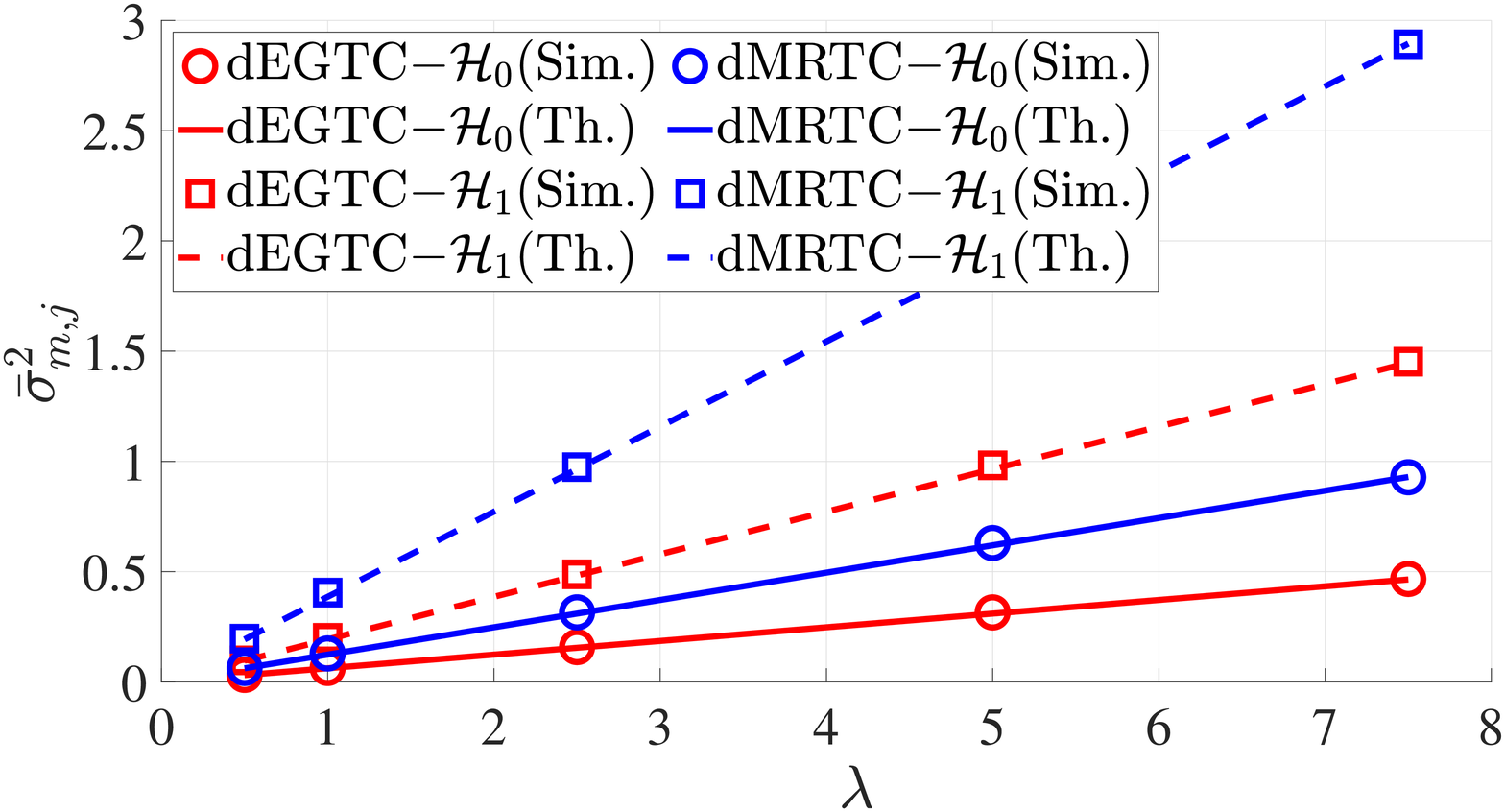}} 
\caption{The mean ($\bar{\mu}_{m,j}$) and variance ($\bar{\sigma}^2_{m,j}$) of the received signal $Y_m$ at the cluster containing the target, under the two hypotheses $\Hyp_j$ for $j=0,1$.}
\label{fig:mean-and-var}
\end{figure*}

\section{Simulation Results and Discussion}
\label{sec:Simulation-Result}
%
%
\begin{table}[]
\centering
\begin{tabular}{|m{17mm}|m{15mm}|m{15mm}|m{15mm}|}
\hline
\textbf{Algorithm Acronym} & \textbf{Fusion rule} & \textbf{Distributed Combining} & \textbf{Distribution} \\ \hline
MOR-dMR-L & Optimal    & dMRTC   & Lognormal \\
\hline
MOR-dMR-N & Optimal    & dMRTC   & Gaussian \\
\hline
MOR-dEG-L & Optimal    & dEGTC   & Lognormal \\
\hline
MOR-dEG-N & Optimal    & dEGTC   & Gaussian \\
\hline
MER-dMR-L & Equal gain & dMRTC   & Lognormal \\
\hline
MER-dMR-N & Equal gain & dMRTC   & Gaussian \\
\hline
MER-dEG-L & Equal gain & dEGTC   & Lognormal \\
\hline
MER-dEG-N & Equal gain & dEGTC   & Gaussian \\
\hline
\end{tabular}
\caption{Distributed detection algorithms.}
\label{tab:DD-Alg}
\end{table}

We simulate a clustered WSN deployed in a $100 \times 100$ unit$^2$ ROI for $5\times 10^4$ Monte Carlo runs. 
The WSN is divided into geographical clusters each having a rectangular shape, in which the CH is located at their centers. The FC is located at the center of the ROI. The target's signal power $P_t = 10$ units. This signal has a path-loss exponent of $\eta = 2$ after a reference distance of $d_0 = 1$ units. The sensing SNR is $\text{SNR}^\mathrm{s} = 12 \text{dB}$. The SNs local detector, $g\left( S(\x_i) \right)$, is chosen to be the matched filter with a local probability of false alarm $P_{fa} = 0.01$. On the communication side, the communication reference distance $r_0 = 1$ unit as well.  Whereas the channel gains are distributed as iid Rayleigh RV with parameter of $\sigma^2_G = 1/\sqrt{2}$. The SNR at the CH is $\SNRchm = 20\,\dB\, \forall m$ and so is the SNR at the FC $\text{SNR}^\mathrm{fc}_m = 20\,\dB\, \forall m$.
The distributed detection algorithms used in the simulations are explained in Tab.~\ref{tab:DD-Alg}. Moreover, dMRTC and dEGTRC are used in a single cluster setting for the sake of comparison.

Fig.\:\ref{fig:mean-and-var} demonstrates the empirical and theoretical mean and variance of the received signal employing dMRTC and dEGTRC under $\H_0$ and $\H_1$ for communication path-loss exponent $\alpha\in\{2,4\}$. 
In this setting, the WSN is divided into $M=4$ clusters and the target is located at $\xb_t=(20,20)$, which is in the southeast cluster. Note however, that under $\H_0$ all clusters' received signal means and variances are equivalent since there is no target present. It is clear that an almost perfect match between the simulated and theoretical mean and variance $\bar{\mu}_{m,j}$ and $\bar{\sigma}^2_{m,j}$ for both dMRTC and dEGTC under $\H_0$ and $\H_1$ for different values of $\lambda$ and for both path-loss exponent conditions. However, under $\alpha = 4$ the mean and variance are considerably lower than the $\alpha =2$ case. This verifies the analytic expressions for mean and variance provided by Props.~\ref{PROP:MEAN}~and~\ref{PROP:VAR}. 

\begin{figure*}[t]
\def\scle{0.18}
\def\wdth{8.5cm}
\def\hght{6cm}
\def\hgt{0.2\textwidth}
\def\wdh{0.45\textwidth}
\centering
\subfloat[ROC at $M = 4,\lambda = 1$, $\SNRchm = 20\,\dB$ and $\alpha = 2$.]{
\includegraphics[width=\wdh,height=\hgt]{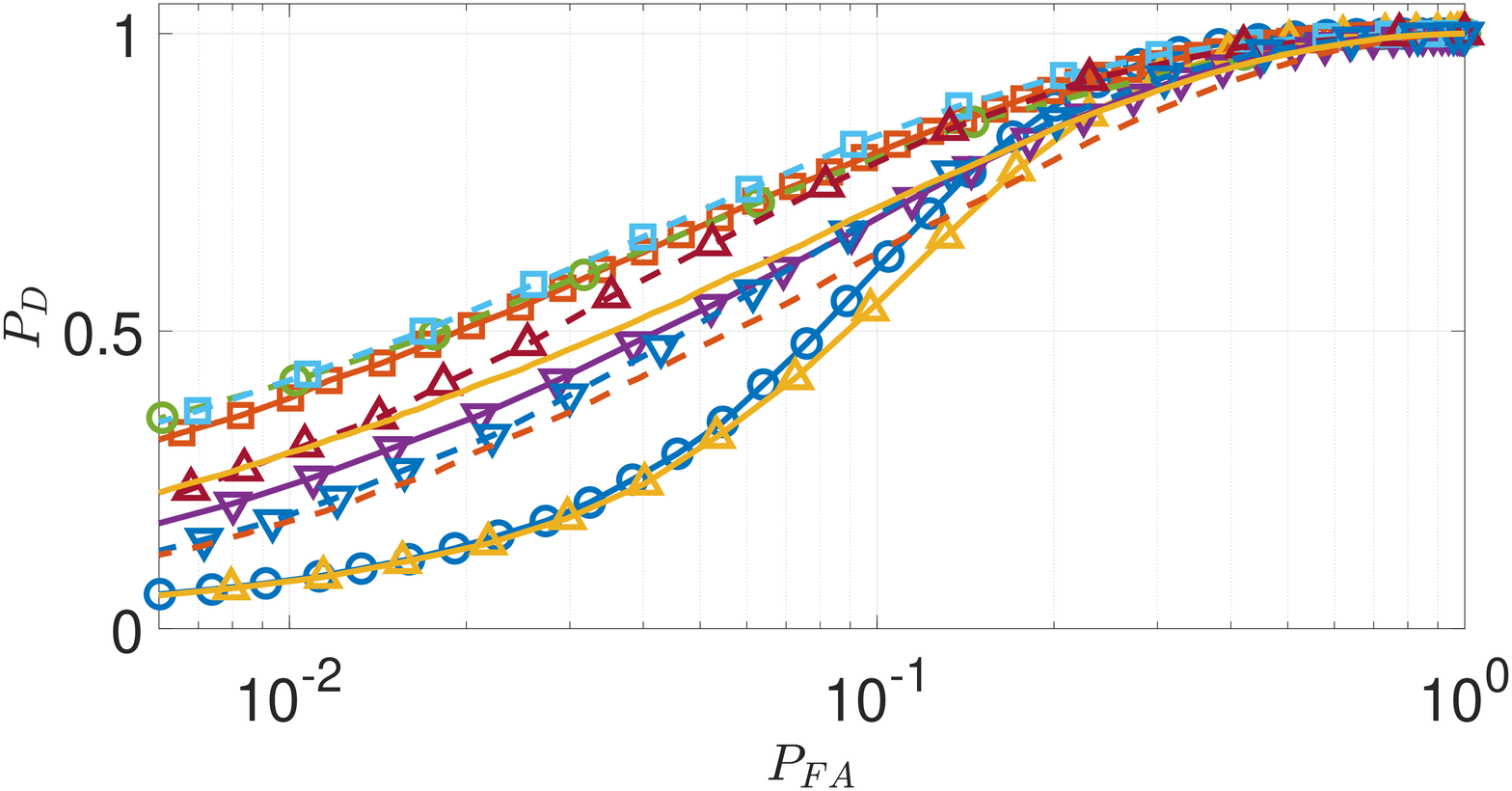}\label{fig:ROC-a}}\hfill
\subfloat[ROC at $M = 9,\lambda = 1$, $\SNRchm = 20\,\dB$ and $\alpha = 4$.]{
\includegraphics[width=\wdh,height=\hgt]{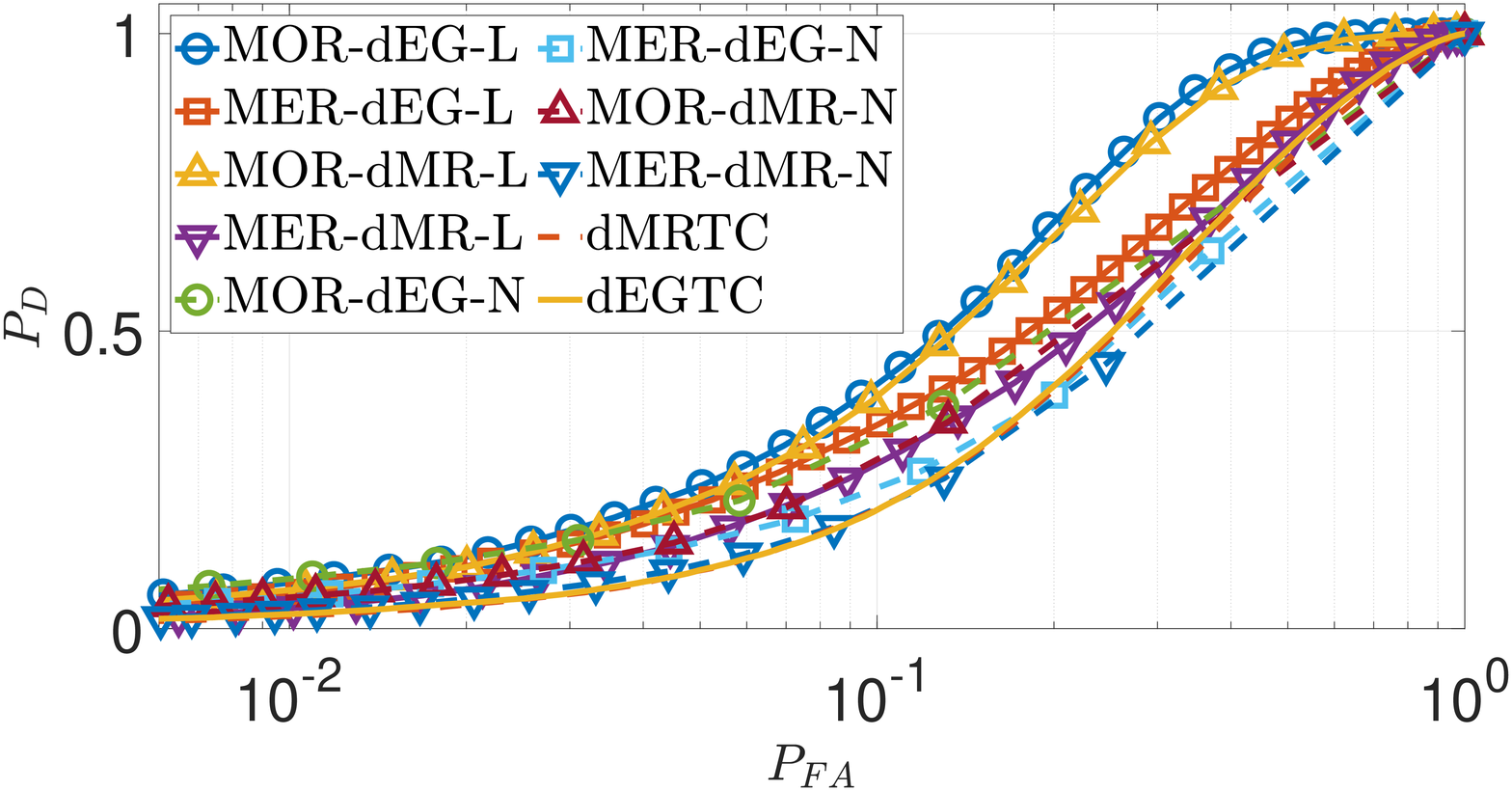}\label{fig:ROC-b}}\hfill
\centering
\subfloat[ROC for $\lambda = 2.5$ at $M = 16$, $\SNRchm = 20\,\dB$ and $\alpha = 2$.]{
\includegraphics[width=\wdh,height=\hgt]{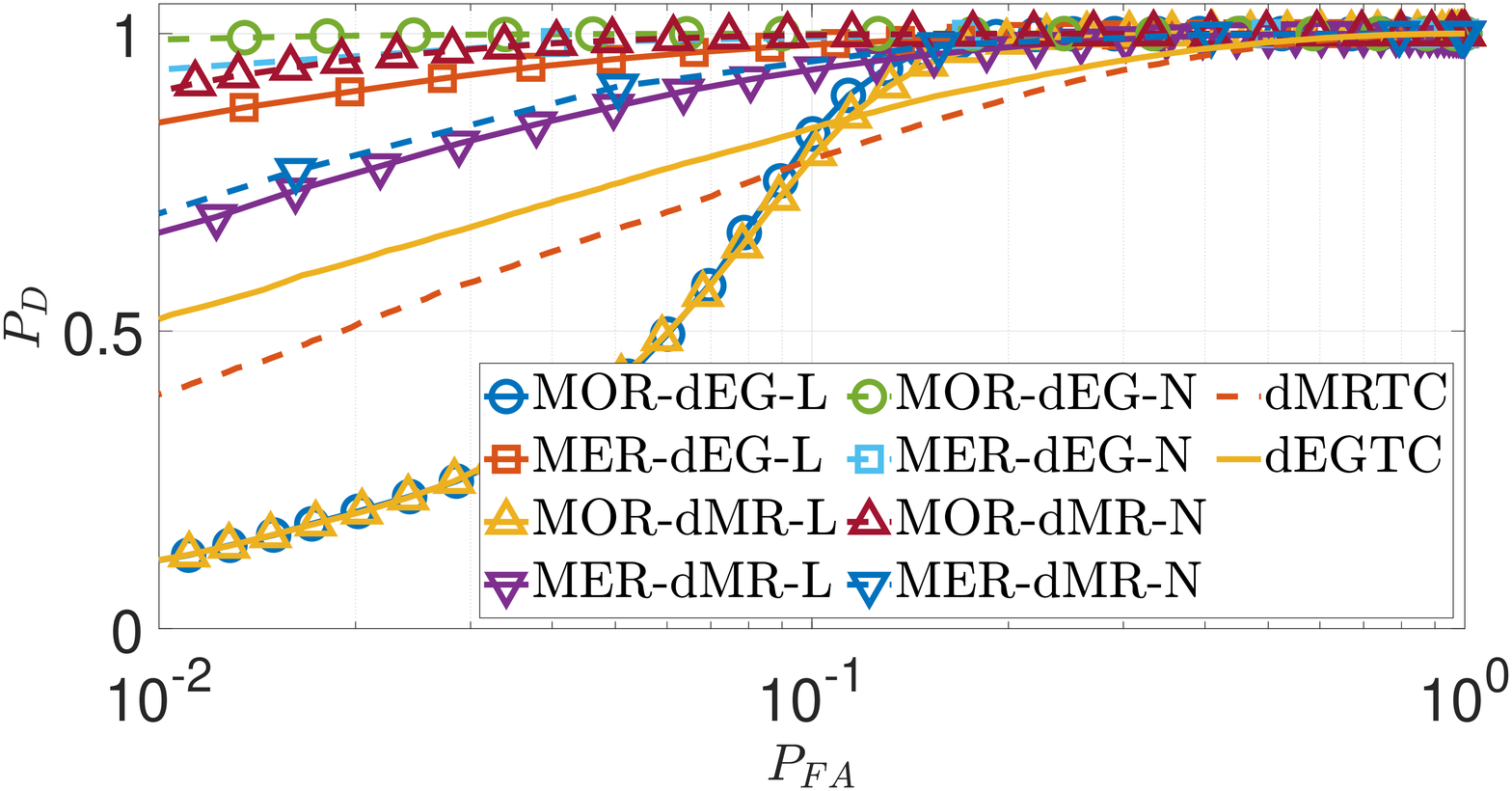}\label{fig:ROC-c}}\hfill
\centering
\subfloat[ROC for $\lambda = 5$ at $M = 16$, $\SNRchm = 20\,\dB$ and $\alpha = 4$.]{
\includegraphics[width=\wdh,height=\hgt]{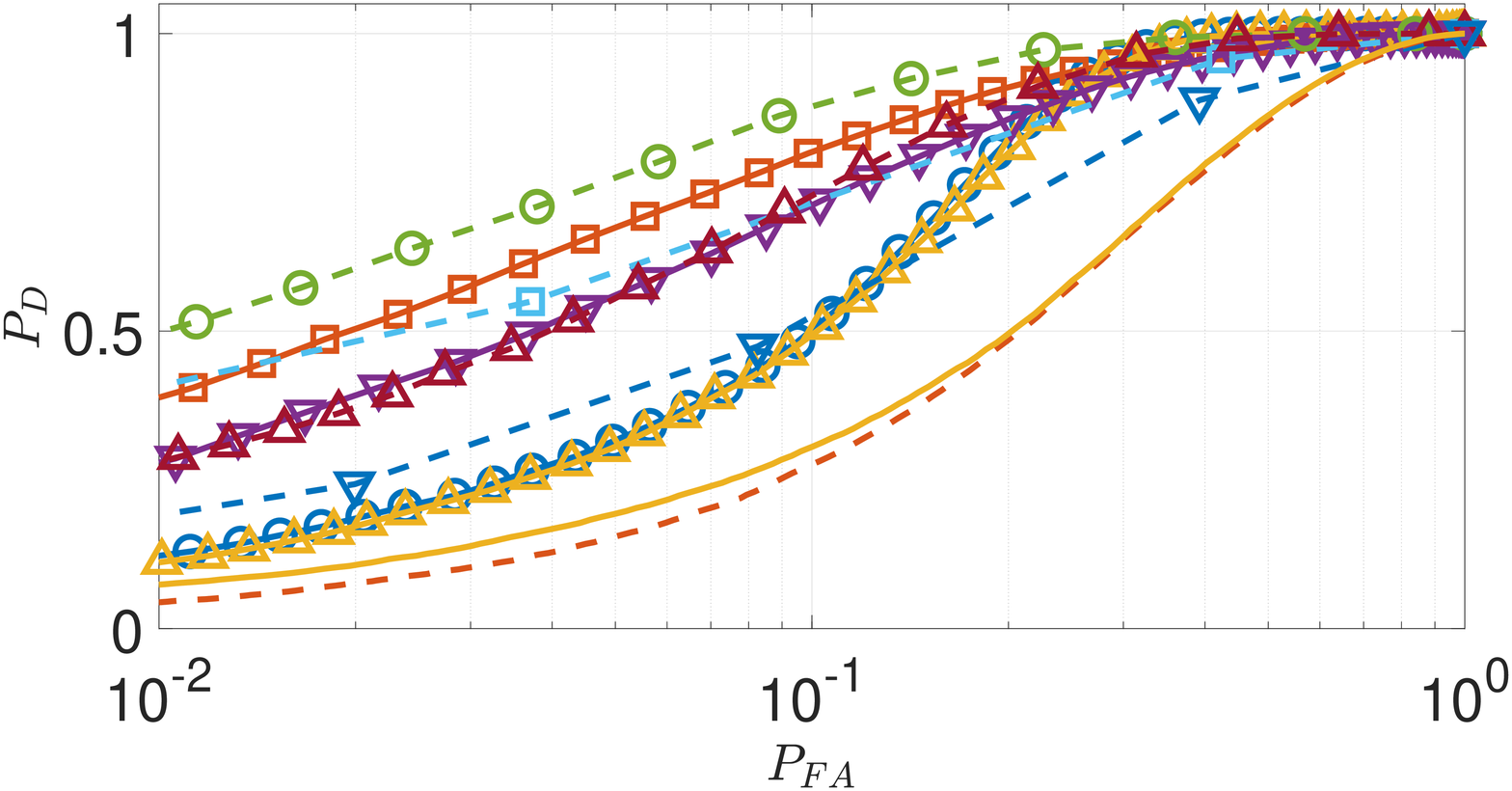}\label{fig:ROC-d}}\hfill
\centering
\subfloat[ROC for $\text{SNR}^{ch}_m = 25\,\text{dB}$ at $M = 16$, $\lambda = 1$ and $\alpha = 2$.]{
\includegraphics[width=\wdh,height=\hgt]{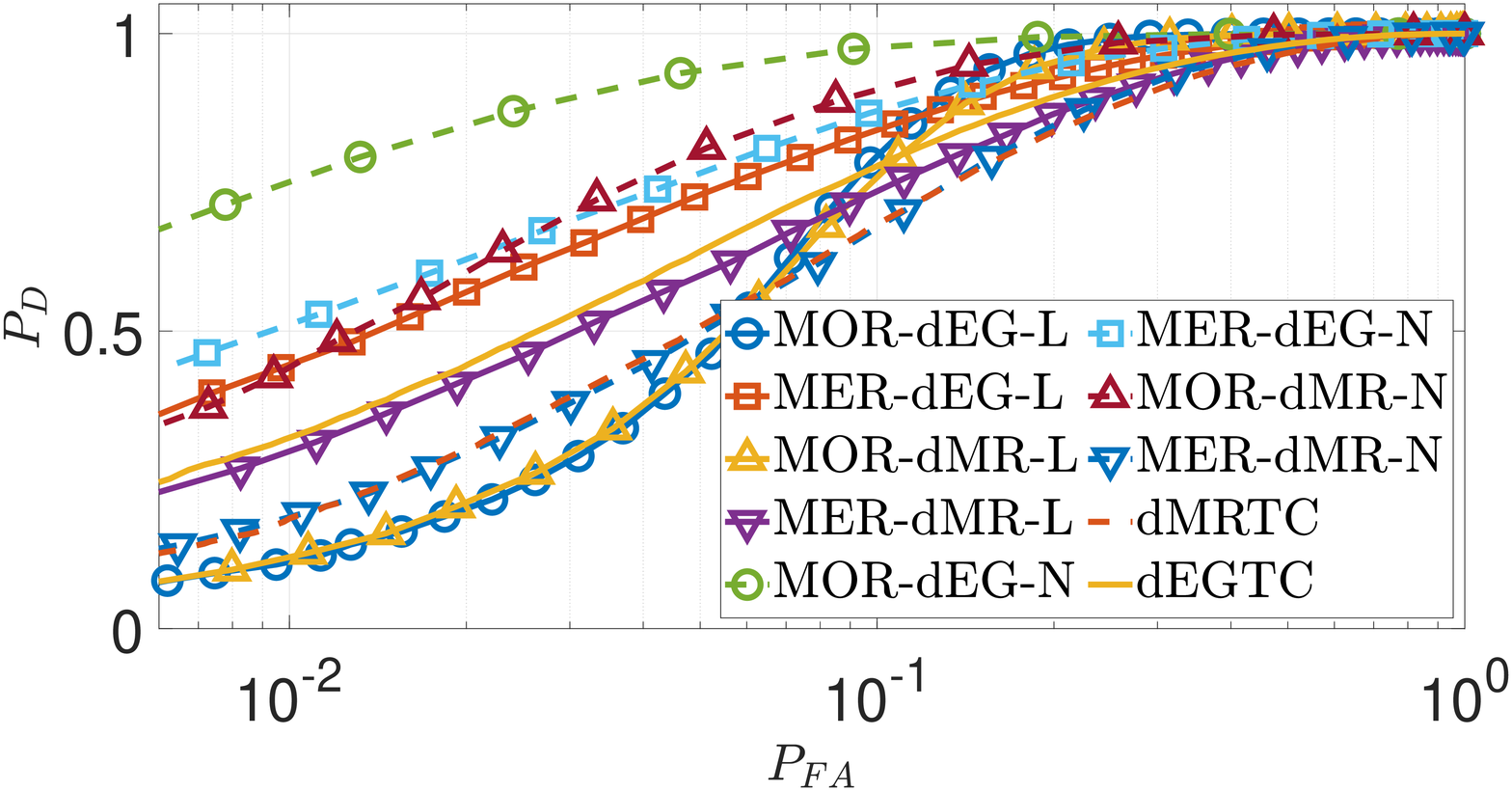}\label{fig:ROC-e}}\hfill
\centering
\subfloat[ROC for $\text{SNR}^{ch}_m = 25\,\text{dB}$ at $M=16$, $\lambda = 1$  and $\alpha = 4$.]{
\includegraphics[width=\wdh,height=\hgt]{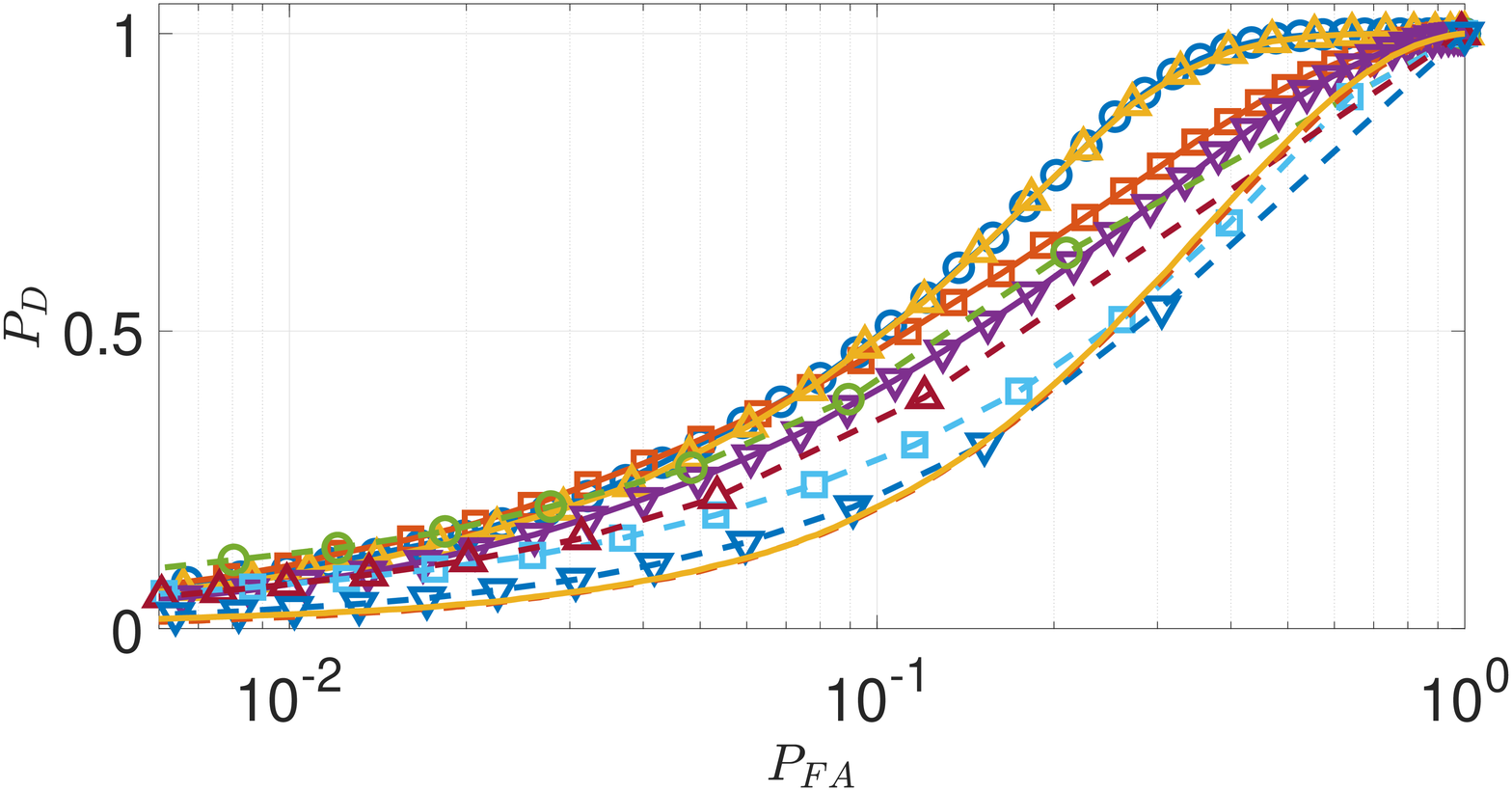}\label{fig:ROC-f}}\hfill
\caption{ROC for various clustered-WSN conditions.}
\label{fig:ROC}
\end{figure*}

Fig.\:\ref{fig:ROC} shows the ROC for the different fusion rules for the same condition parameters as before but with different $M$, $\lambda$ and $\SNRchm$ for both path-loss exponent cases, i.e. $\alpha\in\{2,4\}$. It can be noticed that under certain conditions some algorithms perform better whereas the same algorithms perform worse than others if the conditions are changed. In some instances, the cluster-based algorithms perform worse that the dMRTC and dEGTC rules, which operates in a single cluster. For example, in Fig.\:\ref{fig:ROC-a}, using a single cluster with dEGTC is better than using four clusters using MOR-dEG-L or MOR-dMR-L, while Fig.\:\ref{fig:ROC-d} shows the converse. Hence, we investigate each parameter individually. 

\begin{figure*}[htp] 
\def\scle{0.18}
\def\hgt{0.2\textwidth}
\def\wdh{0.45\textwidth}
\centering
\subfloat[$\alpha = 2$]{
\includegraphics[width=\wdh,height=\hgt]{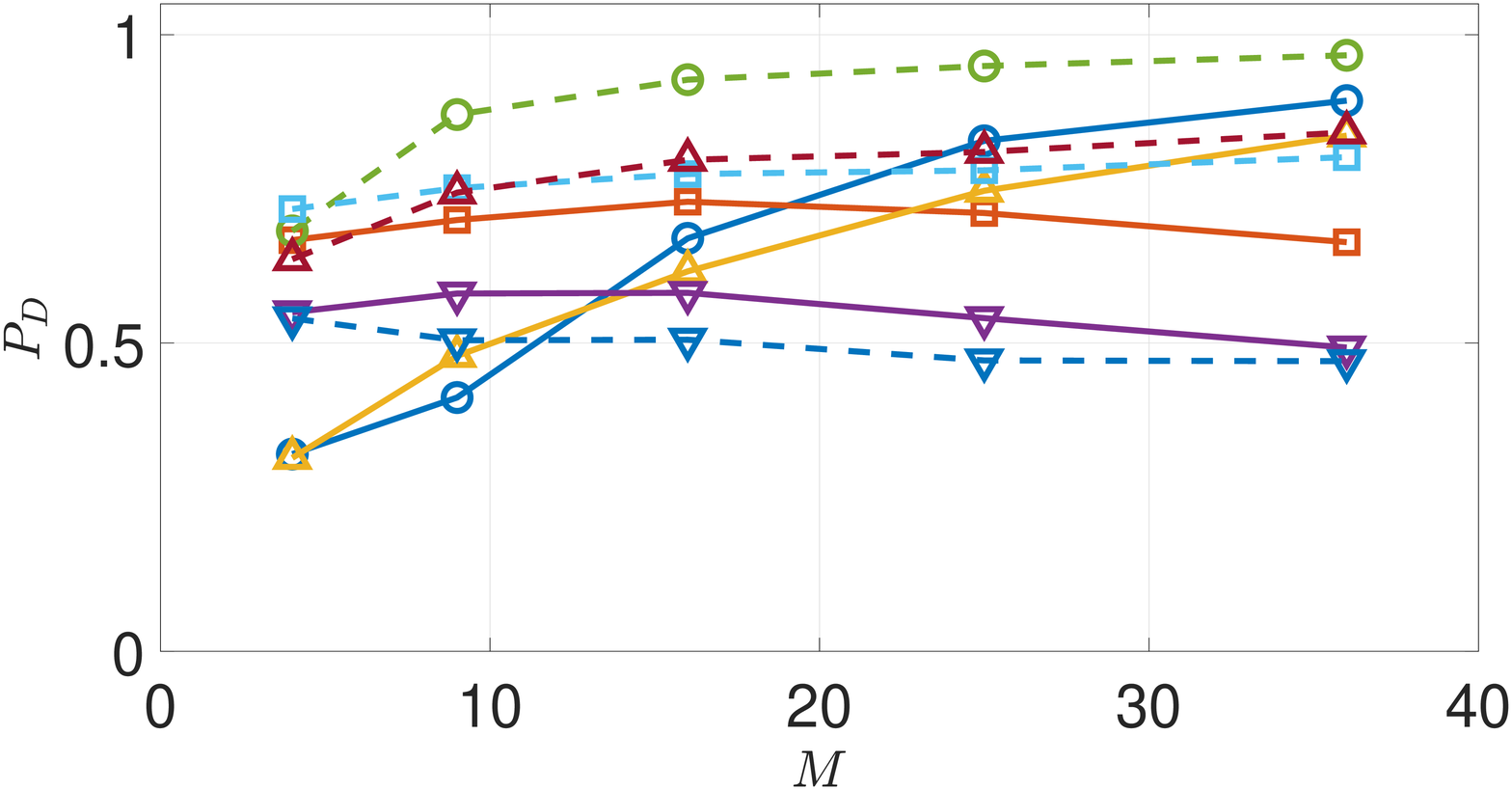}}\hfill
\subfloat[$\alpha = 4$.]{
\includegraphics[width=\wdh,height=\hgtwidth=\wdh,height=\hgt]{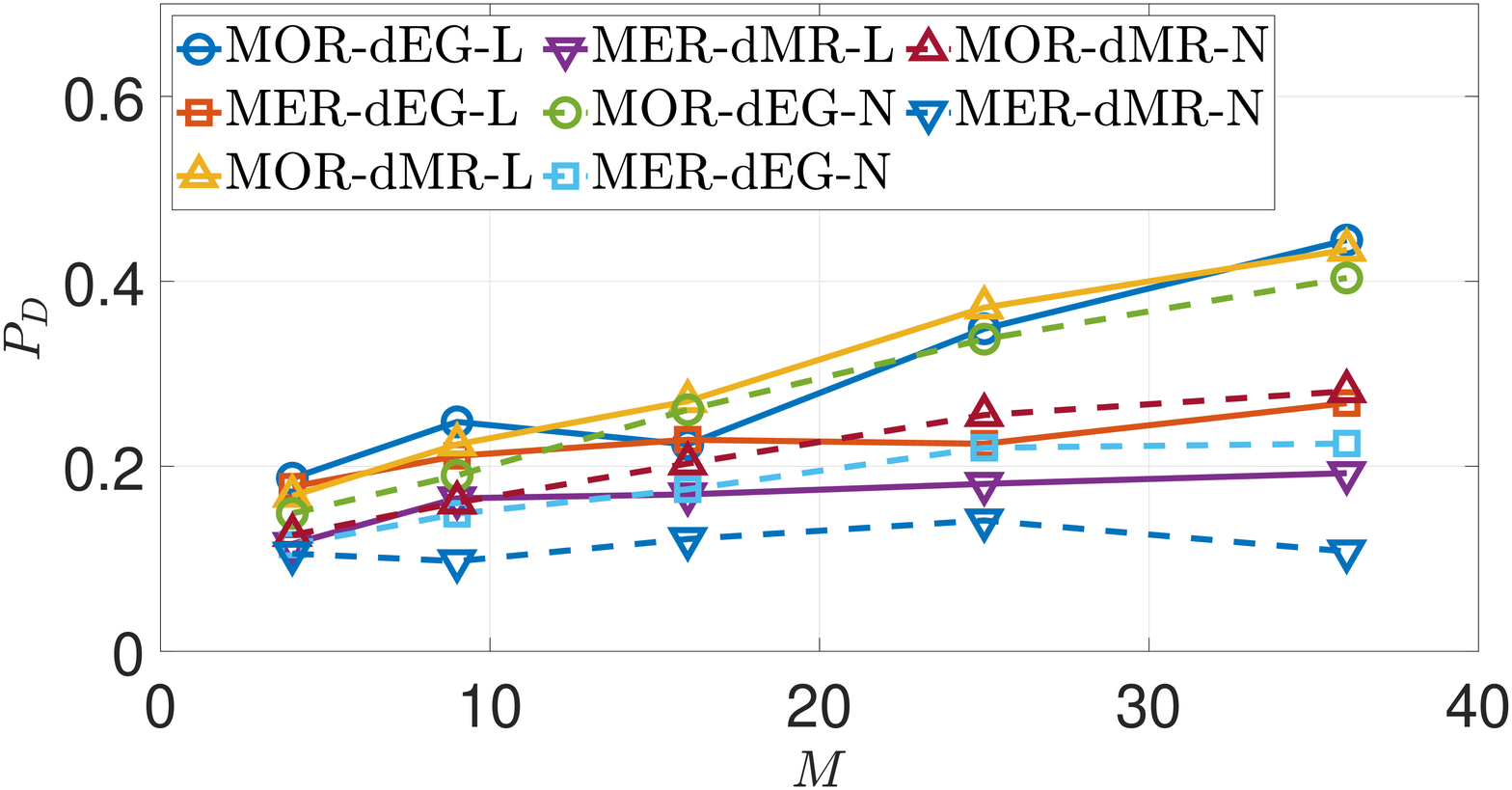}}\hfill
\caption{Probability of detection $P_D$ versus the number of clusters $(M)$ at $P_{FA} = 0.05$, $\SNRchm = 20\,\dB$ and $\lambda = 1$.}
\label{fig:PD-M}
\end{figure*}

Fig.\:\ref{fig:PD-M} shows the detection performance versus varying number of clusters ($M$) when the target is randomly located in the ROI under the above simulation parameters for both $\alpha = 2,4$. The target is randomly located in $85 \times 85$ unit$^2$ area to eliminate the edge effect. In the free-space case, it is clear that the optimal MOR-dEG-N shows superior detection performance, which improves as $M$ increases. The MOR-dMR-N algorithm also improves with $M$, however not as fast as the previous rule. The MER-based algorithms do not improve significantly as $M$ increases. This is due to  not needing any information about the target, so increasing the cluster number does not affect to the performance. In contrast to the MOR algorithms, where having more clusters implies better weighing of the clusters' data and hence better performance. Analogously, the lognormal-derived algorithms follow similar trends but with lower performance. The gap between the normal and lognormal derived algorithms can be explained by the compression effect of the logarithm function in the latter algorithms, which compresses large received signal values at CH in comparison with the Gaussian-derived counterparts leading to a smaller separation between the distributions under $\Hyp_0$ and $\Hyp_1$ and hence worse detection probability. Whereas the dEGTC-dMTRC gap was predicted in Eq. \eqref{eq:Deflection-gain}. As for the $\alpha=4$ case however, the large attenuation reduces the received signals at the CHs forcing the Gaussian-derived algorithms' performance close to or below the lognormal counterparts.

\begin{figure*} 
\def\scle{0.18}
\def\hgt{0.2\textwidth}
\def\wdh{0.45\textwidth}
\centering
\subfloat[$\alpha = 2$]{
\includegraphics[width=\wdh,height=\hgt]{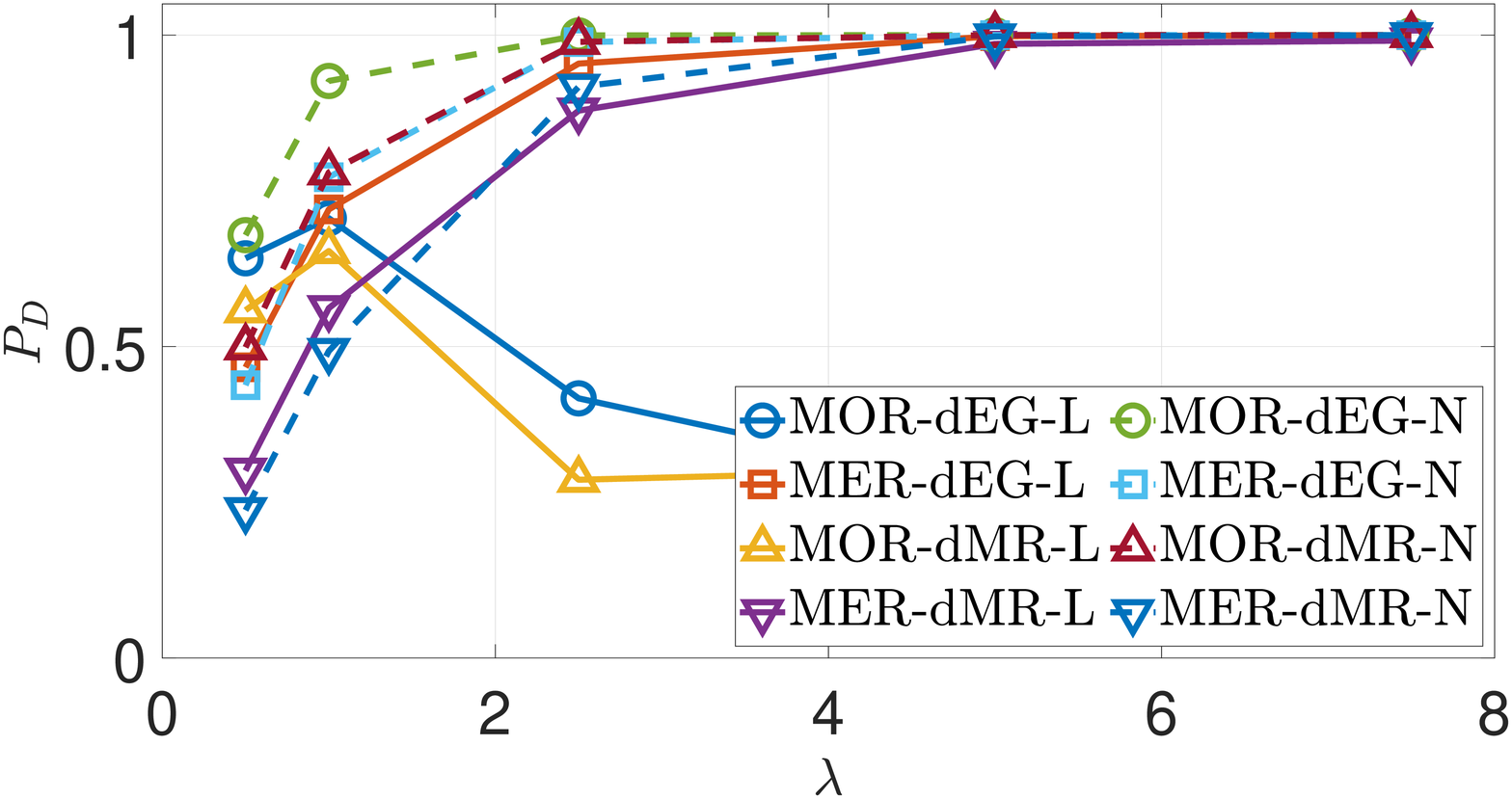}}\hfill
\subfloat[$\alpha = 4$.]{
\includegraphics[width=\wdh,height=\hgt]{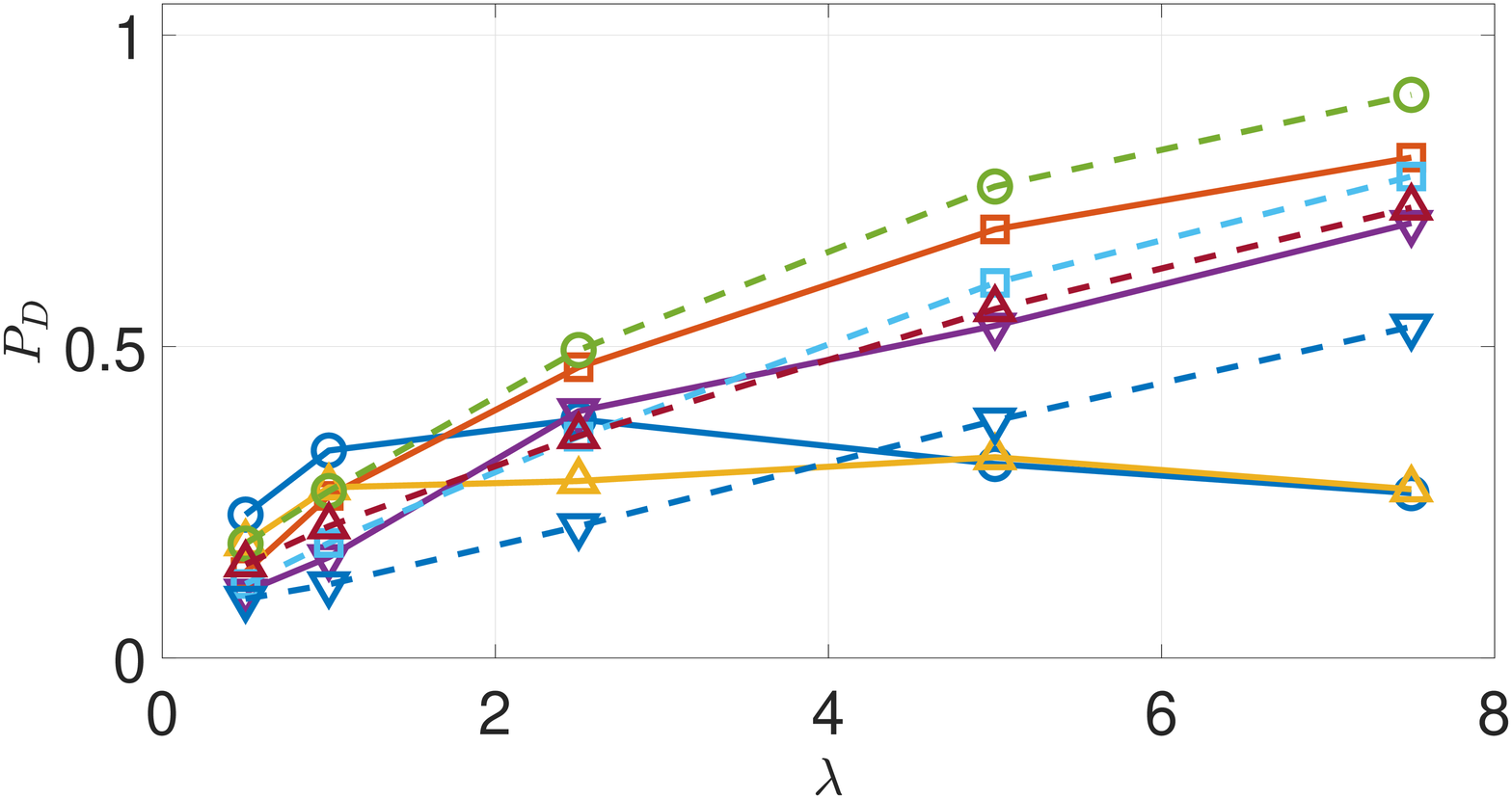}}\hfill
\caption{Probability of detection $P_D$ versus $\lambda$ at $P_{FA} = 0.05$ with $\SNRchm = 20\,\dB$ and  $M= 16$ clusters.}
\label{fig:PD-lambda}
\end{figure*}

Fig.\:\ref{fig:PD-lambda} also shows an improving detection as $\lambda$ increases for both path-loss exponent values (viz. scenarios). This improvement is more pronounced for $\alpha =2$  compared to the $\alpha = 4$ case, due to the least attenuation experienced in the former scenario. However, MOF-dEG-L and MOR-dMR-L algorithms show significant degradation in their performance, which it is more severe in the $\alpha = 2$ case. The compression effect is more pronounced here. In particular, when $\lambda$ increases the CH's received signal does too, but the overall test statistics $\Lambda^L_{MOR}$ is much lower than the $\Lambda^N_{MOR}$. Moreover, the presence of $\hat{a}_m$ reduced the test statistics values compared to the MER-based algorithms, which explains the difference in performance. 

\begin{figure*} 
\def\scle{0.18}
\def\hgt{0.21\textwidth}
\def\wdh{0.45\textwidth}
\centering
\subfloat[$\alpha = 2$]{
\includegraphics[width=\wdh,height=\hgt]{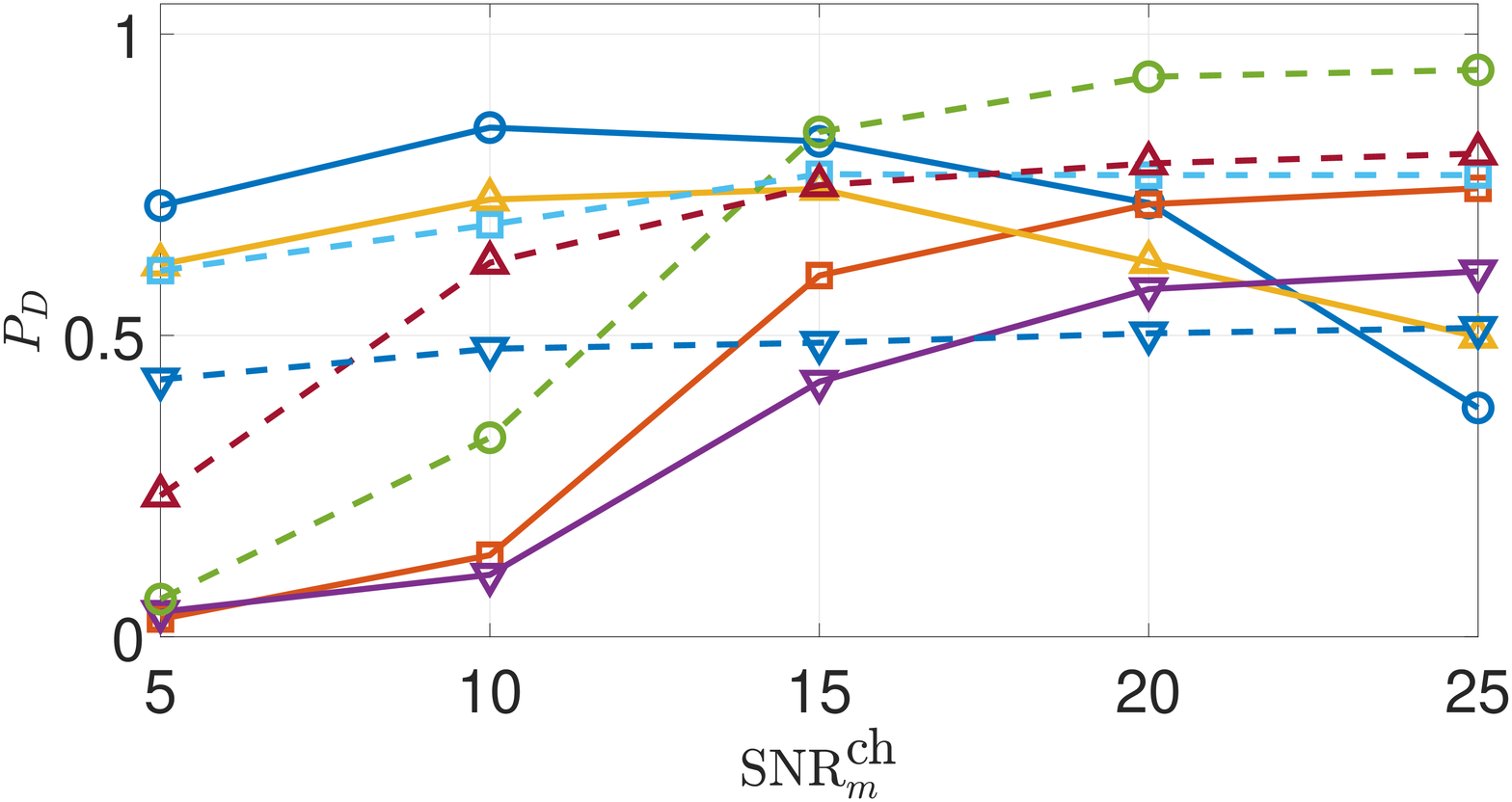}}\hfill
\subfloat[$\alpha = 4$.]{
\includegraphics[width=\wdh,height=\hgt]{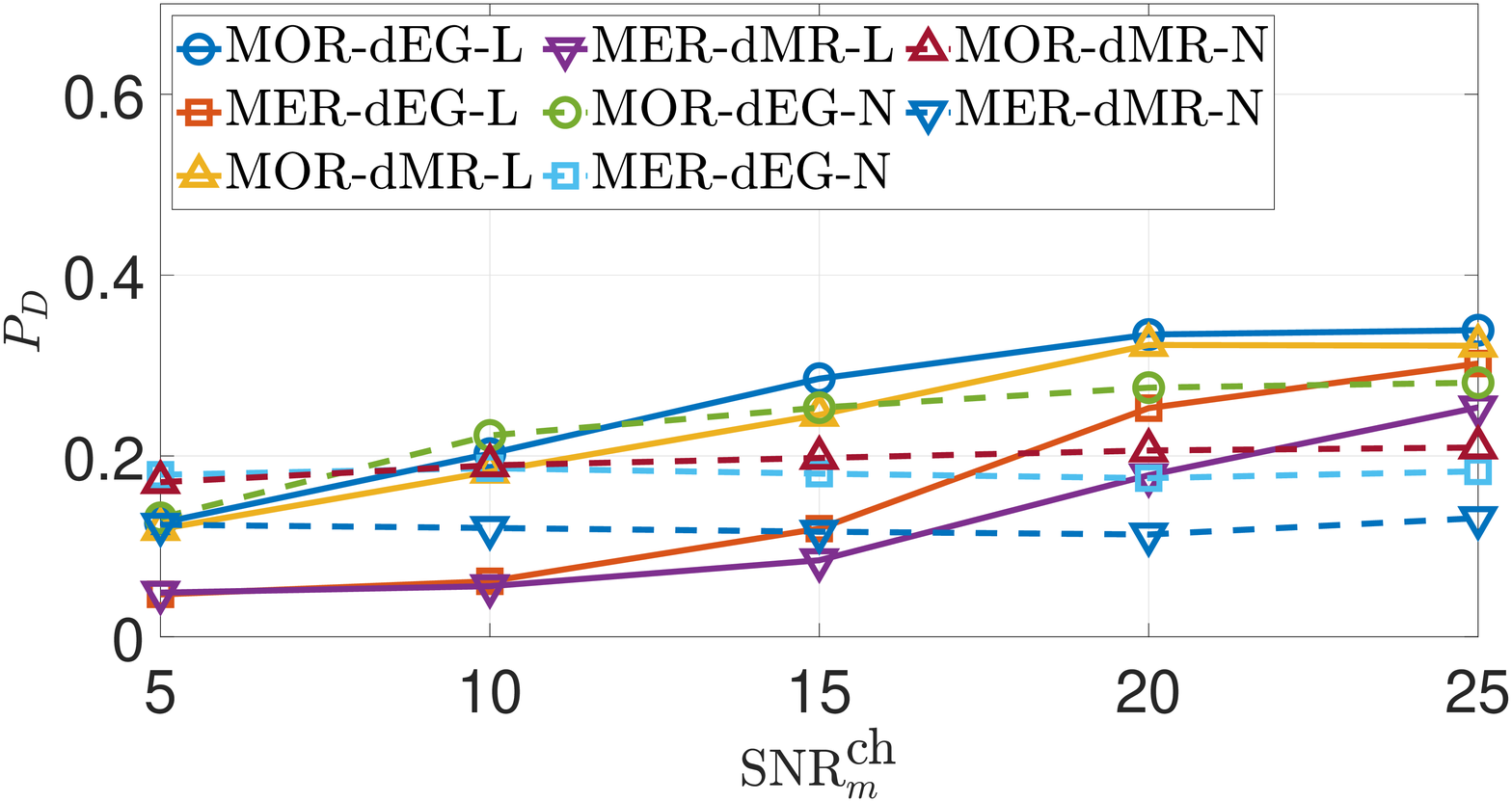}}\hfill
\caption{Probability of detection $P_D$ versus $\SNRchm$ at $P_{FA} = 0.05$ with $\lambda = 1$ and $M= 16$ clusters.}
\label{fig:PD-SNR}
\end{figure*}

The SN-CH channel quality effect on detection performance is demonstrated in Fig.\:\ref{fig:PD-SNR}. It is noticed that the MOR-dEG-L and MOR-dMR-L algorithms show superior performance at relatively low SNRs and then degrades at higher SNRs in the $\alpha =2$ scenario. This is due, again, to the logarithm compression effect, since the CH's received signal have low values and increases when the SNR does. In the $\alpha = 4$ scenario however, the effect of the SNR is nearly neutralized and the MOR-based algorithms show a modest improvement, while the MER counterparts show no such behaviour in general.

\begin{figure*}[t]
\def\scle{0.18}
\def\hgt{0.19\textwidth}
\def\wdh{0.45\textwidth}
\centering
\subfloat[$\alpha = 2$.\label{fig:Prx_nClst-a}]{ 
\includegraphics[width=\wdh,height=\hgt]{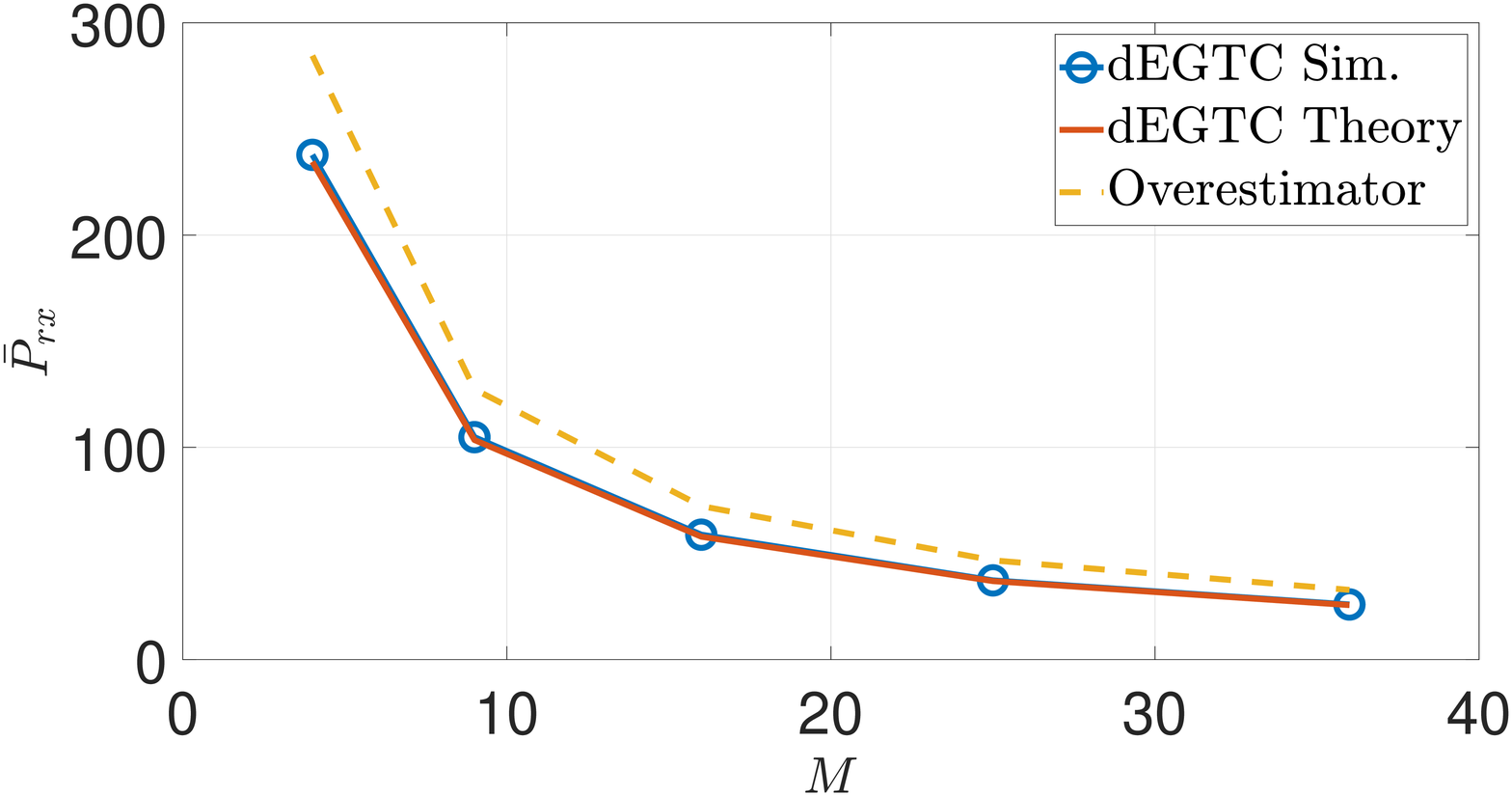}} \hfill
\centering
\subfloat[$\alpha = 4$.\label{fig:Prx_nClst-b}]{ 
\includegraphics[width=\wdh,height=\hgt]{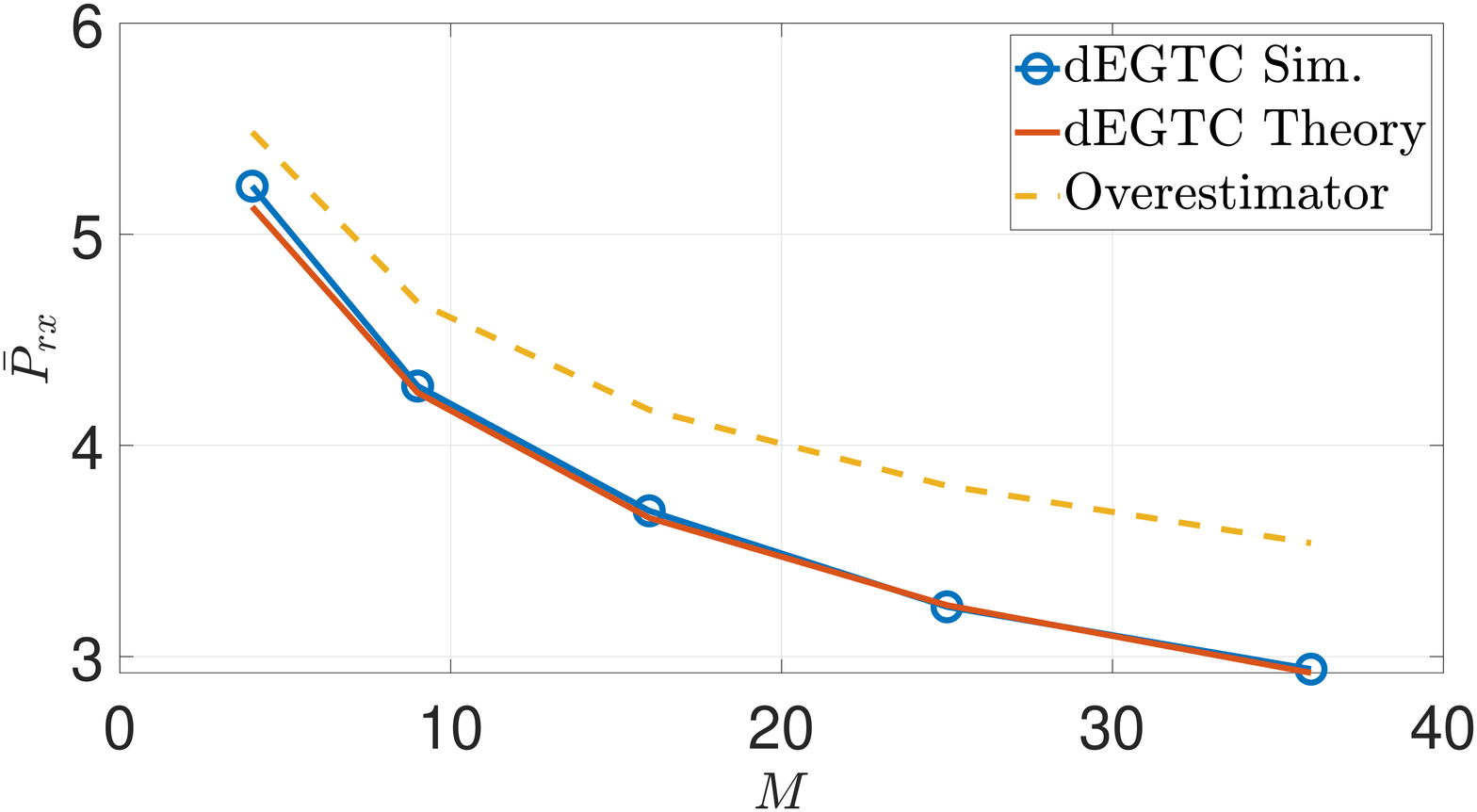}} \hfill
\subfloat[$\alpha = 2$.\label{fig:Prx_nClst-a}]{ 
\includegraphics[width=\wdh,height=\hgt]{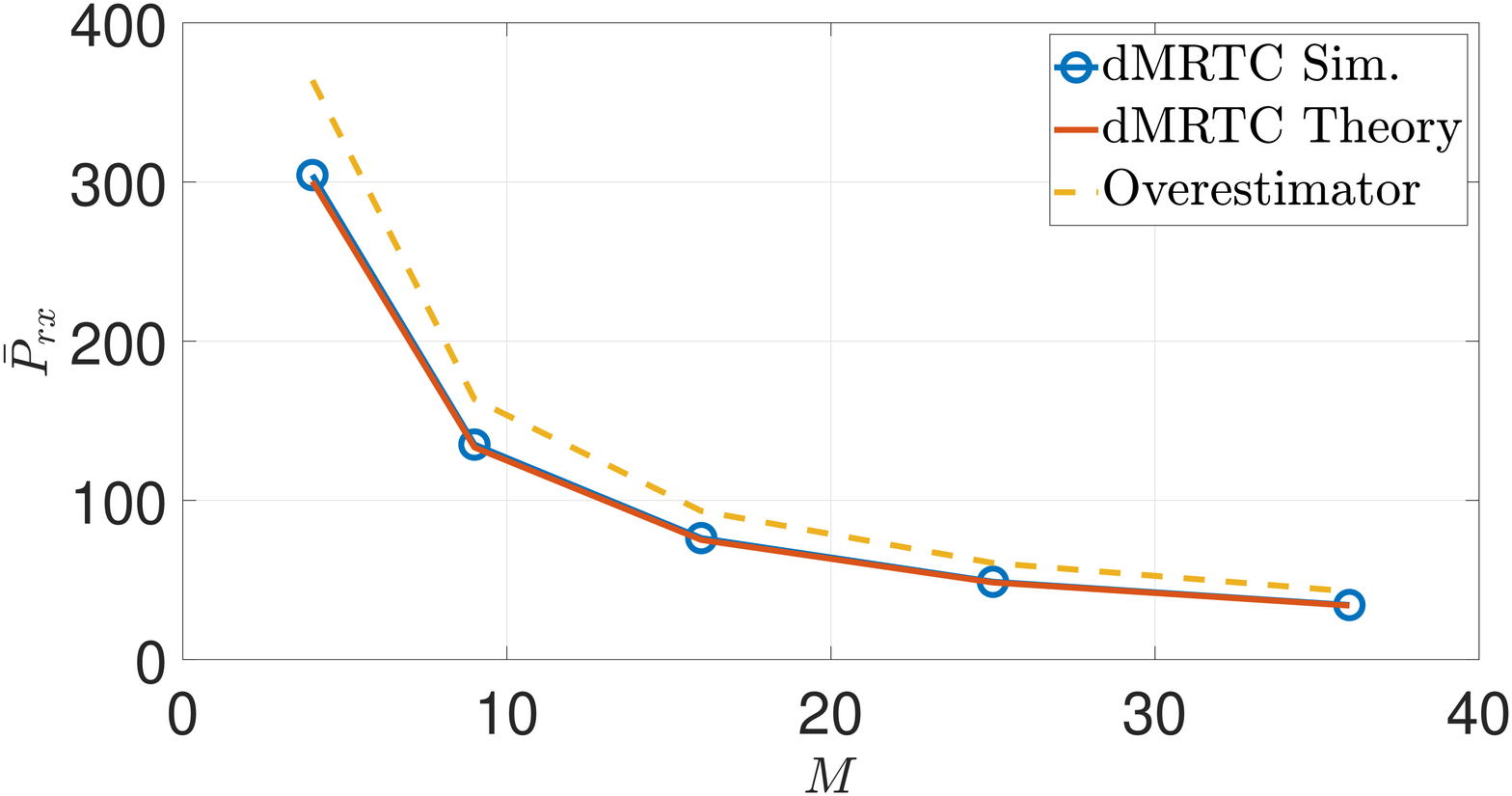}} \hfill
\centering
\subfloat[$\alpha = 4$.\label{fig:Prx_nClst-b}]{ 
\includegraphics[width=\wdh,height=\hgt]{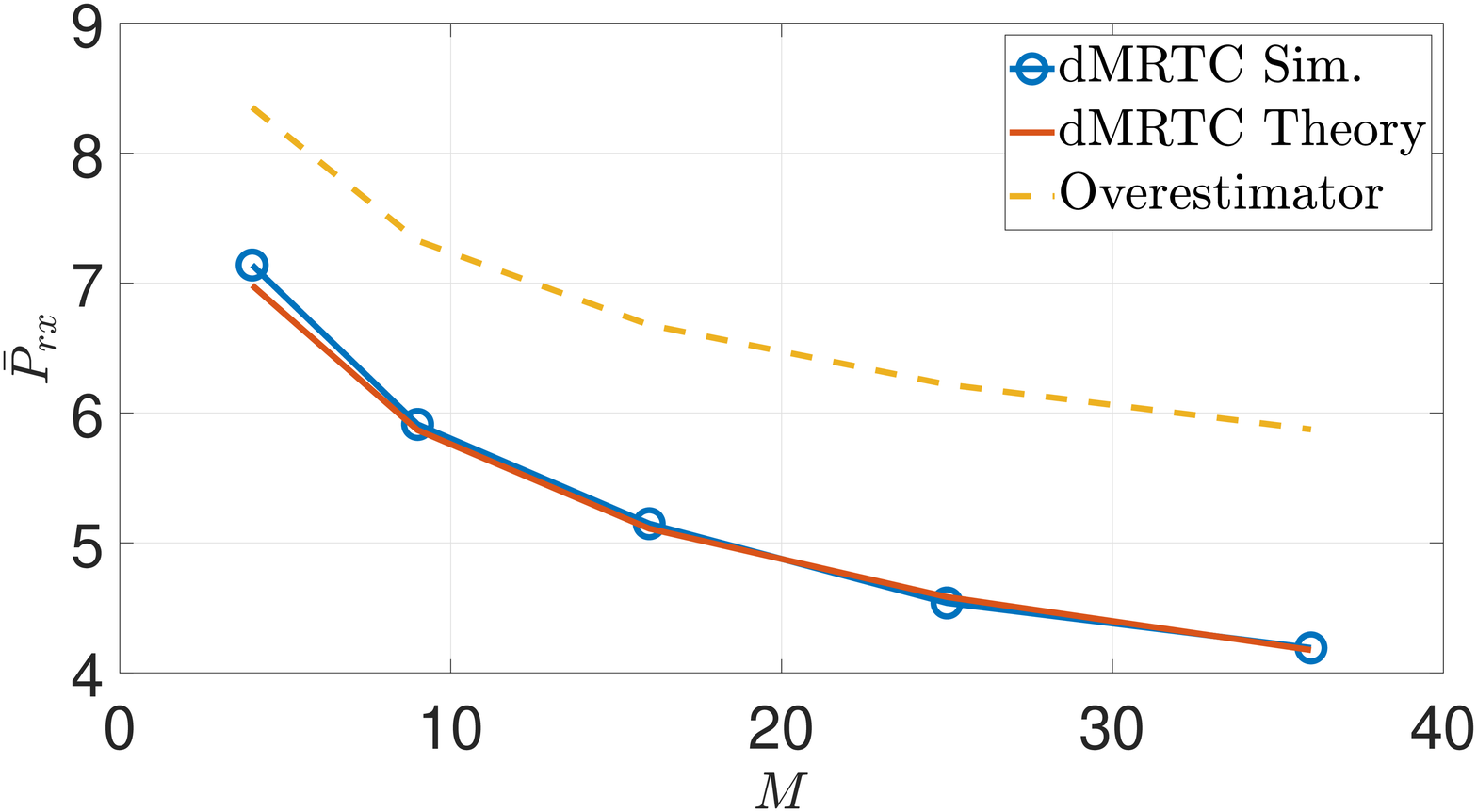}}
\caption{The average received power plotted $\bar{P}_{rx}$ against the number of clusters $M$ at $\lambda =1$ under $\Hyp_0$.}
\label{fig:Prx_nClst} 
\end{figure*}

In order to show the behaviour of the received power at the CH under $\Hyp_0$, we simulate the WSN. Fig.\:\ref{fig:Prx_nClst} shows the averaged received power $\bar{P}_{rx}$ (where the cluster index $m$ was dropped) at the an arbitrary CH. The results show a close match between the simulated power and the exact theoretical value received power for the circular cluster given by Eq. \eqref{eq:Prx_ave} 
\begin{OK}
in Prop. \ref{prop:Prx_ave}. 
\end{OK}
More importantly, it is shown that the received power at the CH decreases as the transmission distance increases when number cluster is increased. Another note here is that the received power in the $\alpha = 4$ is less than the $\alpha = 2$ by an order of magnitude approximately.
Moreover, the power over-estimator proposed in Eq. \eqref{eq:Prx_circ_app} provides a very good estimation for the free-space path-loss case, whereas it provides a reasonable upper bound for ground-reflection case, at least when employing the dEGTC.

%
%
%
%
%
%
%
%
%
%

%% file: 9_conclusions_rev2.tex
\section{Conclusions}
\label{sec:conlcusions}
We investigated distributed detection in clustered WSNs over a shared MAC suffering from Rayleigh fading and additive noise.
To mitigate the effect of the fading channel, two distributed transmit combining methods were proposed, dMRTC and dEGTC.
The statistics of the received signals at the CHs were found via stochastic geometry tools. The latter result was, in turn, used to fit the distribution with log-normal and Gaussian and distributions.
This enabled  deriving a moment matching based optimal fusion rule (MOR) and a simpler moment matching equal gain fusion rule (MER). Interestingly, it has been shown that the dEGTC is better than the dMRTC in terms of the detector's performance.
It has been shown that increasing the number of clusters generally improves the detection performance when knowledge of the target is available. While the MOR-Gaussian based algorithms are better under free-space path-loss with large clusters number, their lognormal counterparts excel in the \emph{ground-reflection} case. Although increasing the SNs deployment density improves the detection performance in general, the MOR lognormal-derived algorithms are better suited for low SNs density and low-SNR WSN scenario, in particular for the free-space propagation. 

The received power at the CHs were derived theoretically and closed-form over-estimator were provided for both propagation cases. An interesting, counter intuitive, result that has been the proved is that the received power at the CH is proportional to $O\left( \lambda^2 R^2 \right)$ for the free-space path loss and to $O\left( \lambda^2 \ln^2 R \right)$ for the ground reflection case.
This shows that the received power scales with increasing cluster size, hence performance improvements can be achieved when the cluster size is increased. From a different point of view, for a given ROI increasing the number of clusters leads to transmission power savings but with the some loss of detection performance (more evident under the ground-reflection scenario). So, the number of clusters can be used to trade-off detection performance for power saving.

\begin{OK}
 As a \emph{future extension} to this work, more realistic SN-CH channel models could be considered, e.g. including large-scale fading (shadowing) effect and fast fading due to potential mobility of SNs~\cite{nurellari2021trajectory}. Furthermore, future work might account for~\emph{heterogeneous SNs} and their effect on distributed detection fusion rules.
 Finally, the challenging generalization of the present study to enumerating multiple intruders (e.g. via model-order selection techniques) is also seen as an interesting venue for related applications such as collaborative spectrum sensing~\cite{quan2008collaborative}.
\end{OK}

%% file: 9b_Appendices_rev5.tex
%
\appendices

%
%
\section{Proof of Proposition \ref{PROP:VAR}}\label{sec:App-A}
Using the total variance identity, the conditional variance of $\bar{Y}_m$ can be written as
\begin{gather}
\sigma_{m,j}^{2}=\var_{\Phi_{m}}\left(\E_{H}\left[\left.\sumPhim\frac{\sqrt{P_{tx}}f(H_{m.i})\I}{\Xixm}\right|\Phi_{m};\H_{j}\right]\right)\nonumber \\
+\E_{\Phi_{m}}\left[\left.\var_{H}\left(\sumPhim\frac{\sqrt{P_{tx}}f(H_{m.i})\I}{\Xixm}\right)\right|\Phi_{m};\H_{j}\right]. \nonumber \\
\end{gather}

Next, due to the i.i.d property%
\footnote{In particular, it holds $\var\left(\sum_{i=1}^{N}a_{i}S_{i}\right)=\sum_{i=1}^{N}a_{i}^{2}\var(S_{i})$.}
of the $f(H_i)$'s the variance simplifies as
\begin{gather}
\sigma_{m,j}^{2}=P_{tx}\,\E_{H}^{2}\left[f(H)\right]\var_{\Phi_{m}}\left(\sumPhim\frac{\I}{\Xixm};\H_{j}\right)\nonumber \\
+P_{tx}\,\var_{H}\left(f(H)\right)\E_{\Phi_{m}}\left[\sumPhim\frac{\I}{\Xixmsq};\H_{j}\right].\label{eq:var}
\end{gather}
We use Campbell's theorem on the first term above as follows. For a given $g(\xb)\geq 0$, we can write $\var\left( \sumPhi g(\xb) \right) = \lambda \int g^2(x) d\xb = \E \left[ \sumPhi g^2(\xb) \right] $.
Setting $\ensuremath{g(\cdot)=I(\cdot)\,/\left\Vert (\cdot)-\mathbf{x}_{m}\right\Vert ^{\frac{\alpha}{2}}}$, the first term in Eq. Eq. \eqref{eq:var} becomes
\begin{gather}
P_{tx}\,\E_{H}^{2}\left[f(H)\right]\,\var_{\Phi_{m}}\left(\sumPhim\frac{\I}{\Xixm};\H_{j}\right)\nonumber \\
=P_{tx}\,\E_{H}^{2}\left[f(H)\right]\,\E_{\Phi_{m}}\left[\sumPhim\frac{\I}{\Xixmsq};\H_{j}\right].
\end{gather}

Then using the variance identity $\left( E[S^2]=\var(S)+E^2[S] \right)$, Eq.~\eqref{eq:var} further reduces to
\begin{equation}
\sigma_{m,j}^{2}=\E\left[f^{2}(H)\right]P_{tx}\E_{\Phi_{m}}\left[\sumPhim\frac{\I}{\Xixmsq};\H_{j}\right].
\end{equation}

Finally, applying Campbell's theorem yields Eq. \eqref{eq:var_j}.

%
%
\section{Proof of Theorem \ref{THM:MOFR-N}}\label{sec:App-MOR-N-Proof}
We assume that the distribution of  $Z_m|\mathcal{H}_j$ is Gaussian with mean $\mu_{m,j}$ and variance $\sigma^2_{m,j}$ for $j=0,1$. 
The
\begin{OK}
Neyman-Pearson
\end{OK}
likelihood ratio then is 
\begin{equation}
    \Lambda_1 = \prod^M_{m=1} \left(  \frac{\sigma_{m,0}}{\sigma_{m,1}} \right) \frac{\exp\left( -\frac{1}{2\sigma^2_{m,1}}\left(z_m - \mu_{j,1} \right)^2\right)}{\exp\left( - \frac{1}{2\sigma^2_{m,0}}\left(z_m - \mu_{j,0} \right)^2\right)}. 
\end{equation}

The corresponding log-likelihood ratio $\Lambda_2 = \log \Lambda_1$ is
\begin{gather}
\Lambda_2  = \SumM\ln\frac{p\left(z_{m}|\,\H_{1}\right)}{p\left(z_{m}|\,\H_{0}\right)}=\SumM\,\ln\left(\frac{\sighmj{0}}{\sighmj{1}}\right)\nonumber \\ 
-\dfrac{1}{2\sigma^2_{m,1}} \left(z_m - \mu_{j,1} \right)^2 +\dfrac{1}{2\sigma^2_{m,0}} \left(z_m - \mu_{j,0} \right)^2 
\end{gather}.

Neglecting the terms independent of $z_m$ and expanding and arranging the terms in a quadratic form w.r.t. $z_m$ we get
\begin{equation}
    \Lambda_3 = \SumM a_m z^2_m +  b_m z_m  + c_m 
    \label{eq:Lambda-3-Quad}
\end{equation}
where
\begin{align}
    a_m & =  \dfrac{1}{2\sigma^2_{m,0}} - \dfrac{1}{2\sigma^2_{m,1}}, \\
    b_m & = \dfrac{\mu_{m,1}}{\sigma^2_{m,1}} - \dfrac{\mu_{m,0}}{\sigma^2_{m,0}},\\
    c_m & = \dfrac{\mu^2_{m,0}}{2\sigma^2_{m,0}} - \dfrac{\mu^2_{m,1}}{2\sigma^2_{m,1}}.
\end{align}
Looking at Eq. \eqref{eq:Lambda-3-Quad}, we recognize a quadratic expression in $z_m$, hence completing the square we can write 
\begin{align}
\Lambda_3 & = \SumM a_m \left(z_m + \frac{b_m}{2a_m} \right)^2 + c_m - \frac{b^2_m}{4a^2_m} \\
\Lambda_4 & = \SumM a_m \left(z_m + \frac{b_m}{2a_m} \right)^2 
\end{align}
where the constant term was ignored in $\Lambda_4$ and 
\begin{eqnarray}
d_{m} & \triangleq & \frac{b_{m}}{2a_{m}} =  \frac{\sigma^2_{m,1}\mu_{m,0}-\sigma^2_{m,0}\mu_{m,1}}{\sigma^2_{m,1}-\sigma^2_{m,0}}
\end{eqnarray}
giving the fusion rule in Eq. \eqref{eq:NP-FR}.
%
%
\section{Proof of Theorem \ref{THM:MOR-L}}\label{sec:App--MOR-L-Proof}
First, we note that the (fitted) distribution of $Z_m|\mathcal{H}_j$ is 
\begin{gather}
p(z_{m}|\mathcal{H}_{j})=\frac{1}{\sqrt{2\pi} z_{m}\sighmj{j}}\exp\left(-\frac{\left(\ln \vert z_{m} \vert -\muhmj{j} \right)^{2}}{2\sighSqmj{j}}\right)
\end{gather}
where the $\vert \cdot \vert$ is used to avoid singularities, the mean ($\muhmj{j}$) and variance ($\sighSqmj{j}$) of the RV's natural logarithm are directly
 related to the mean ($\mu_{m,j}$) and the variance ($\sigma^2_{m,j}$) as:
\begin{eqnarray}
\muhmj{j} &=& \ln \left( \frac{\mu^2_{m,j}}{\sqrt{\sigma^2_{m,j}+\mu^2_{m,j}}} \right), \label{eq:mu_a} \\
\sighmj{j} &=& \ln \left( 1 + \frac{\sigma^2_{m,j}}{\mu^2_{m,j}} \right). \label{eq:sigma2_a}
\end{eqnarray}
The log-likelihood ratio is then 
\begin{gather}
\Lambda_5 = \SumM\ln\frac{p\left(z_{m}|\,\H_{1}\right)}{p\left(z_{m}|\,\H_{0}\right)}=\SumM\,\ln\left(\frac{\sighmj{0}}{\sighmj{1}}\right)\nonumber \\
+\SumM\left\{ \frac{\left(\ln \vert z_{m} \vert -\muhmj{0}\right)^{2}}{2\SigmaAM{0}}-\frac{\left(\ln \vert z_{m} \vert -\muhmj{1}\right)^{2}}{2\SigmaAM{1}}\right\}. \label{eq:LLR-L-1} 
\end{gather}

Neglecting terms independent of $z_m$, expanding and re-arranging the terms in \eqref{eq:LLR-L-1}, yields
\begin{gather}
\Lambda_6 = \SumM\,\hat{a}_{m}\,\ln^{2}\vert z_{m} \vert +\hat{b}_{m}\,\ln \vert z_{m} \vert +\hat{c}_{m}\label{eq:Lambda_LLR_1}
\end{gather}
where we have exploited the following auxiliary definitions $\hat{a}_m  \triangleq (2\sighSqmj{0})^{-1} - (2\sighSqmj{1})^{-1}$, $\hat{b}_m  \triangleq \hat{\mu}_{m,1} (\sighSqmj{0})^{-1} - \hat{\mu}_{m,0} (\sighSqmj{1})^{-1} $ and $\hat{c}_m \triangleq \hat{\mu}^2_{m,1} (2\sighSqmj{0})^{-1} - \hat{\mu}^2_{m,0} (2\sighSqmj{1})^{-1}$.

Following the same proof method adopted in Appendix \ref{sec:App-MOR-N-Proof} leads to the fusion rule in Eq. \eqref{eq:MOF-L}. 